# Development and Experimental Evaluation of Grey-Box Models for Application in Model Predictive Control of a Microscale Polygeneration System

**Parantapa Sawant*[1]. Adrian Bürger[2,3]. Minh Dang Doan[2]. Clemens Felsmann[4]. Jens Pfafferott[1]**

[1]Institute of Energy Systems Technology (INES), Offenburg University of Applied Sciences, Offenburg
(Tel: 0049-781- 205-4696; E-Mail: parantapa.sawant(a)hs-offenburg.de)
[2]Systems Control and Optimization Laboratory, Department of Microsystems Engineering (IMTEK), University of Freiburg, Freiburg
[3]Faculty of Management Science and Engineering, Karlsruhe University of Applied Sciences, Karlsruhe
[4]Chair of Building Systems Engineering and Heat Supply, Technische Universität Dresden, Dresden

Abstract: With the need for optimisation based supervisory controllers for complex energy systems, comes the need for reduced order system models representing not only the non-linear characteristics of the components, but also certain unknown process dynamics like their internal control logic. We present in this paper an extensive literature study of existing methods and a rational modelling procedure based on the grey-box methodology that satisfies the necessary characteristics for models to be applied in an economic-MPC of a real-world polygeneration system at the Offenburg University of Applied Sciences. The engineering application of the models and their fitting coefficients are shared in this paper. Finally, the models are evaluated against experimental data and the efficacy of the methodology is discussed based on quantitative and qualitative arguments.

## 1. INTRODUCTION

Applying Model Predictive Control (MPC) for the optimal scheduling of a decentralised polygeneration system has shown promising results for their energy-efficient, sector-coupled (power-to-heat or gas-to-electricity) and grid-reactive operation (Cho et al., 2014; Gu et al., 2014; Jradi and Riffat, 2014; Liu et al., 2014; Wu and Wang, 2006). Researchers have quantified potential operational cost savings of 2 to 6% for buildings with storages (Cole et al., 2012) and 29% for medium scale trigeneration systems (Cho et al., 2009b). Potential energy savings between 15% to 28% for a building heating system (Široký et al., 2011), 24.5% for a building cooling system (Ma et al., 2009), and 8.5% for a medium scale trigeneration system (Chandan et al., 2012) are also reported in the literature.

However, the common consensus in the research community regarding gaps in status of MPC application for building technology is the lack of engineering demonstration projects on the supply side and research that implements HVAC components readily available on the market (Bruni et al., 2015; Cho et al., 2014; Dagdougui et al., 2012; Jradi and Riffat, 2014). This brings with it challenges to develop experimentally evaluated models that are not only able to simulate a wide range of operating conditions and represent the non-linear dynamics of the components with sufficient accuracy, but also are simple enough for application in optimisation-based control algorithms.

With these premises, the contribution of this paper is the application of the grey-box methodology to develop optimisation-suitable system models and experimentally evaluating their plausibility for implementation in MPC of a complex energy system. This includes a novel formulation for the model of a continuously differentiable stratified water storage tank that is adaptable to its constructional features as well as the validation of a silica-gel based adsorption chilling machine model against experimental data. A practical conclusion to the different model evaluation metrics applied in the HVAC field is also made in the final section of this paper.

In the next section, we briefly describe the polygeneration plant set-up and the challenges faced for modelling the system, followed by a summary of the extensive literature research and the proposed techniques. In the fourth section, the individual component models are described in detail. Then, experimental and simulation results from different functional tests are compared and analysed.

## 2. PROBLEM DESCRIPTION

At the Institute of Energy Systems Technology (INES) at Offenburg University of Applied Sciences, a microscale polygeneration plant has been installed using standard industrial components (Sawant and Pfafferott, 2017). The main specifications of the primary parts are given in *Table 1.* In the near future, it is planned to operate this plant in a grid-supportive and cost efficient manner by applying optimal control.

Table 1 Important specifications of the main components

| Components | Abbr. | Specification |
|---|---|---|
| Adsorption Chiller | AdCM | 12 $kW_{th}$ cooling capacity max.; 0.65 max coefficient of performance |
| Combined Heat and Power Unit | CHP | 5 $kW_{el}$; 10 $kW_{th}$; 59% $\eta_{th}$; 30% $\eta_{el}$; fuel oil |
| Outdoor Coil | OC | 0.9 $kW_{el}$ at 480 RPM |
| Reversible Heat Pump | RevHP | 12.9 $kW_{th}$ (cooling capacity nom.); 16.7 $kW_{th}$ (heating capacity nom.); 3.75 $kW_{el}$ (power input nom.) |
| Thermal Loads | Load_H / Load_C | Water cooled thermostats (10 $kW_{th}$ cooling and 18 $kW_{th}$ heating capacity) and approx. 20 m² of thermally activated building systems (ca. 3 $kW_{th}$) |
| Thermal Energy Storage | HTES / CTES | 1500 L / 1450 L |

The formulation of an optimisation problem that controls different aspects of this entire plant such as volume flows, mixing temperatures, fan-speed and component switches etc. would be very complex because of the number of variables and constraints that must be considered. A possible solution for such engineering systems is to use a multi-level architecture or a cascaded control strategy to reduce the number of states and decision variables in the core optimisation problem (Lefort et al., 2013; Picasso et al., 2010; Scattolini, 2009). Additionally, this architecture facilitates the inclusion of the stable and efficient internal controllers of the individual components and their hydraulic circuits. It also facilitates the implementation of widely accepted Data Acquisition and Control frameworks (DAQ) in the field of building automation and control. The resulting architecture for the Automation and Control System (ACS) of the plant is shown in *Fig. 1* and consists of three levels: field level, automation level and management level.

At the field level, the present operating states (temperatures, volume flow etc.) are measured via sensors and the positioning of valves and switching of components is done via actuators. The sensor data are transmitted to the automation level as feedback signal to indicate the settings of monitoring equipment. The HVAC components on the field level are controlled directly by positioning signals from the lower level controllers acting on mixing valves, cooling tower fans and pumps. Thermal loads are generated using a Hardware in the Loop (H-i-L) set-up on the field level mentioned in *Table 1*.

At the automation level, plant data are monitored and while analog signals (electrical resistance, voltage etc.) are converted into digital signals before transmission.

At the management level a supervisory controller in the form of a moving horizon MPC will be programmed in the project. An algorithmic differentiation based optimisation framework in CasADi interfaced to Python® (Andersson et al., 2019) will be used for this purpose. The algorithm will solve an optimisation problem at each sampling time based on the component models, forecast data (weather, load and energy prices) and actual system states. The optimal control signal only for the first time-step will be applied to the field level through the automation level during the following sampling interval. At the next time-step, a new optimal control problem using current measurements of system states will be solved over a shifted horizon.

In this design, we aim to anticipate the plant production and consumption over a 24-hour prediction horizon using mathematical models to calculate an optimal dispatch schedule that minimises the cost of operation. Consequently, the ability of the models to simulate with adequate accuracy and speed over the length of this prediction horizon greatly influences the quality (stability & practicality) of the controller. This leads to certain challenges in terms of developing the models.

To overcome these challenges we first expressed them as desired characteristics of the models to make them appropriate for application in a supervisory optimal control problem that works in tandem with a field level controller. These characteristics are shown in *Fig. 2*.

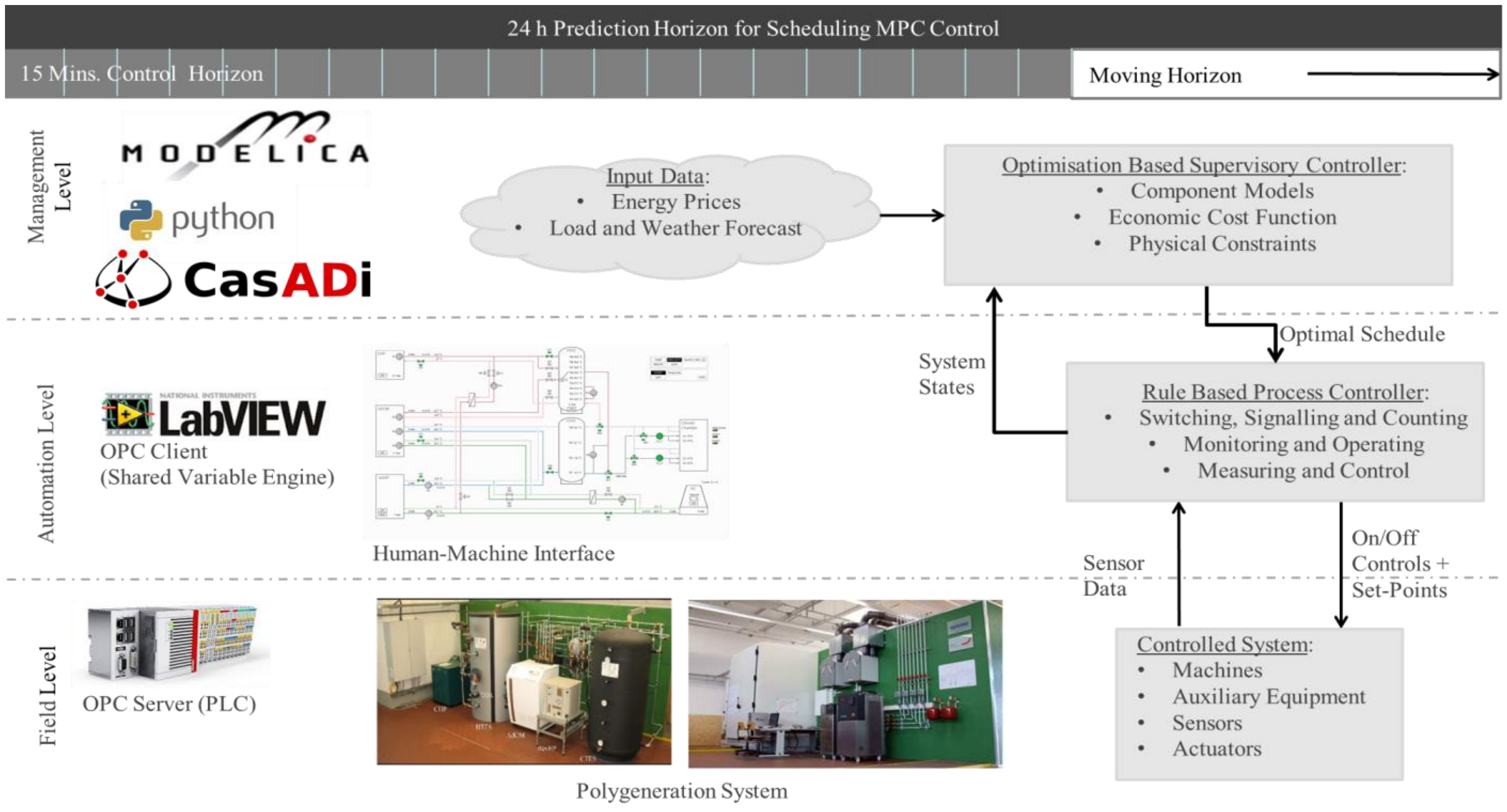

Fig. 1 A hierarchical architecture for the Automation and Control System (ACS) of the INES polygeneration lab.

visualised in a LabVIEW® based Human-Machine Interface (HMI). A rule-based controller can independently control the plant using conventional control strategies if the optimal control at the management level fails. The binary signals are processed directly and transmitted to the management level,

*Dynamic characteristics*: When operating with actual components a switch from one operating point to another often has a dynamic effect on the states of the system and this should be included in the models for improving the controllability of the system (Zhou et al., 2013). Thus, in an MPC loop for a thermal system, if the components display slow dynamic behaviour extending far in excess of the sampling time interval, then this behaviour must be simulated accordingly.

*Part-load behaviour*: Likewise, on the field level, if the components have an internal control logic that uses low-level controllers to improve their performance under part-load conditions or due to safety considerations, then this information should be available to the supervisory controller through the models of the system.

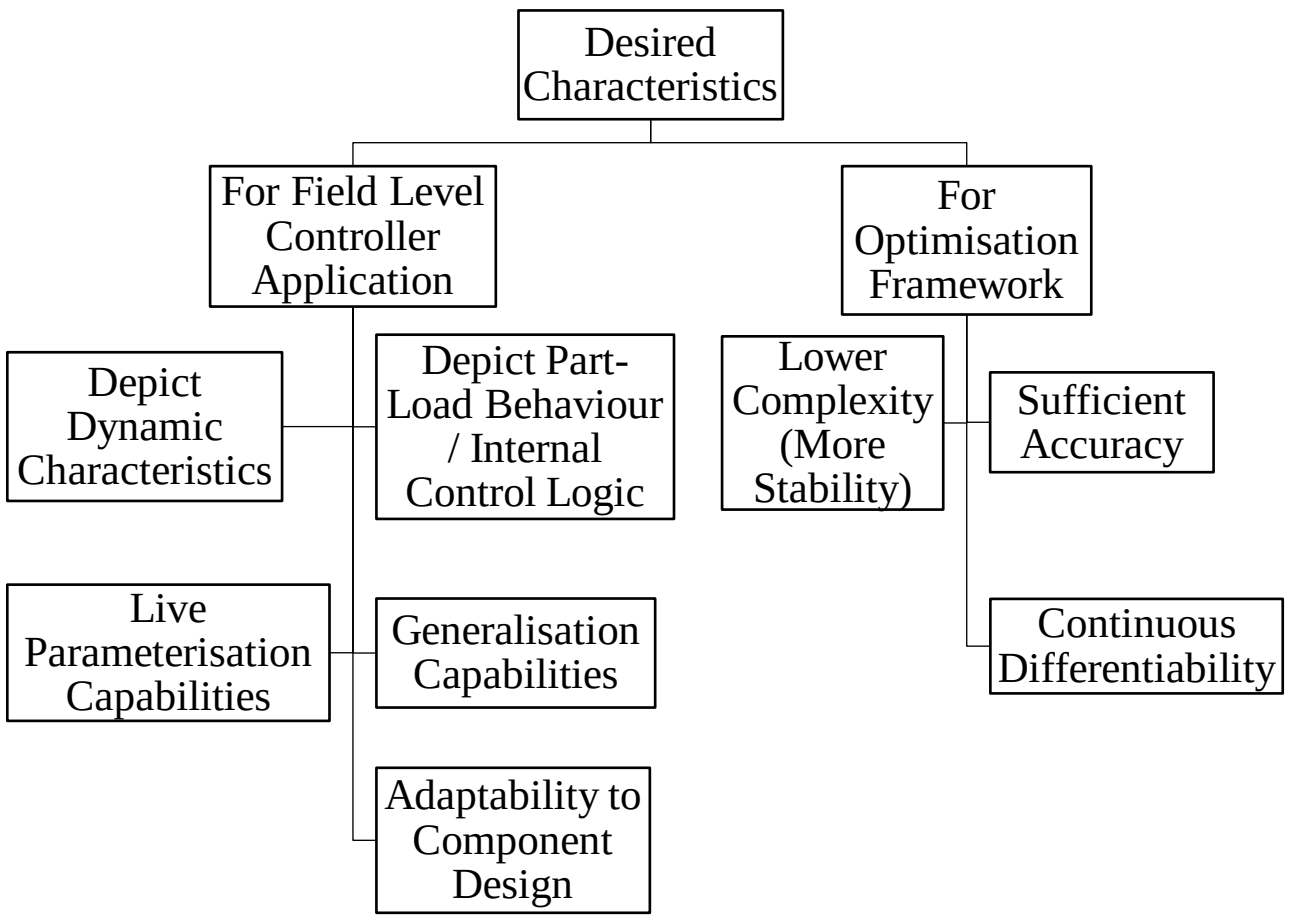

Fig. 2 Desired characteristics in system models.

*Live parameterisation*: If the MPC approach is to be integrated in a retrofit scenario, then the rationale behind modelling the components should permit for live parameterisation. For example, regression-based models should use data that is readily collected in standard industrial practices and do not need specialised instrumentation.

*Generalisation*: Likewise, if the MPC approach is to be integrated in a green-field scenario, then the rationale behind modelling the components should permit for generalisation capabilities. For example, regression-based models should use data for fitting the coefficients that is readily available from standard industrial data sheets of the components or can be collected during the commissioning phase of the plant.

*Component design*: In certain cases, the constructional design of the component influences its performance and its interoperability within the entire system directly. A concrete example of this in building energy systems is the construction of the storage tank. The height at which water enters or leaves the tank will depend on the hydraulic connections to the tank or the type of heat transfer to the water in the tank will depend on the type of heat exchanger in the tank. Thus, the models of such components should have the ability to adapt to the type of design.

*Complexity*: In their paper on modelling and optimisation of a trigeneration system, Chandan *et al.*, 2012 pointed out very clearly the unsuitability of detailed HVAC simulation models for direct use in an MPC structure due to their large computational times and other associated challenges (Chandan et al., 2012). For example, for each new state the size of the entire optimisation problem increases by a factor of the total number of sampling intervals over the entire forecast horizon. Thus, the number of states and parameters in the models for the system components should be limited to achieving the relevant results from the perspective of a thermo-economic optimisation.

*Accuracy*: Although model reduction might compromise on the generalisation or accuracy of the models, only a sufficient accuracy of the models is needed for MPC of thermal systems that typically demonstrate slow dynamics (Lefort et al., 2013; Široký et al., 2011). This is based on the advantage of the moving horizon optimisation framework that gives means to adjust the control and react to disturbances due to the update of system information after every sampling time.

*Differentiability*: The models need to be continuously differentiable since their application is in gradient-based optimisation methods (Biegler, 2010; Wächter and Biegler, 2006).

In the next section, we present the literature that was reviewed to identify existing methods and models that could either completely or partly fulfil the above characteristics. Then, the grey-box modelling method and related techniques used in this work are briefly explained.

## 3. METHODOLOGY

Extensive literature research was done to identify the state-of-the-art in modelling techniques applied in the field of HVAC simulation and control. As shown in *Table 2* the models were sorted as per different factors to identify existing models that could be adapted for our application. It was noticed that many simulation models for standard HVAC components were available and a further in-depth qualitative analysis of some selected models was done. This is summarised in *Table 3*.

We determined that although some optimal scheduling problems are already implemented, they often use simplified linear power flow models and do not consider the hydraulic circuit temperatures or do not represent the behaviour of the internal controller of the components. The existing models that satisfy these requirements are either physics based models with a component level simulation focus making them very complex, or have rule-based controllers making them not continuously differentiable.

Nevertheless, this analysis revealed important features of different models that were then partly adapted in our approach. These adaptations are described in more detail in the corresponding component-model in the next section.

Based on the qualitative analysis, on the results of previous functional tests (Sawant and Pfafferott, 2015) and on the guidelines given in other literature reviews (Afram and Janabi-Sharifi, 2014; Trčka and Hensen, 2010) we decided to follow the grey-box method to model our system. In grey-box modelling, the first law of thermodynamics and the principle of mass and energy balance is applied for developing the mathematical structure of the models and any missing variables or unknown physical processes are quantified through data fitting methods. This method is a compromise between completely physics based white-box models and data-driven black-box models and can provide good generalization capabilities while maintaining a level of accuracy better than physics based models (Afram and Janabi-Sharifi, 2015a; Bohlin, 1994; Sohlberg, 2003). Grey-box models are also robust to disturbances, have auto-tuning capabilities, and need fewer assumptions to set-up. This is an advantage over data-driven algorithms like artificial neural networks for developing black-box models with promising results but limitations on generalisation capabilities and less robustness to disturbances (Afram and Janabi-Sharifi, 2015b).

Table 2 Classification of literature for HVAC systems modelling.

| | Modelling Class / Methodology | Main Outputs | Size/Complexity of Model | Objective of Study/Model and Validation | Reference(Years Ascending) |
|---|---|---|---|---|---|
| **AdCM:**<br><br>Silica gel / water | • Nonlinear dynamic white-box models<br>• LDF kinetic equation for adsorption and desorption rate<br>• Pressure and enthalpy based mass and energy balance<br>• Resistance-capacitance heat exchanger model and mass and energy balance | CC, CT and COP | • > 20 parameters<br>• > 4 states<br>• 2 curve fits | To study the effects of circuit temperatures, switching time and cycle time on the AdCM performance<br>Validation : Visual, quantitative (APE, RMS error, NSD) | (Sakoda and Suzuki, 1984), (Saha et al., 1995), (Chua et al., 1999), (Li and Wu, 2009), (Schicktanz and Núñez, 2009) |
| - | • Nonlinear dynamic white-box models<br>• DAE formulations<br>• Density and temperature based mass and energy balance | CC, CT and COP | • 22 parameters<br>• 10 differential states<br>• 2 algebraic states | Application in an optimal control problem to calculate the optimal cycle times of an AdCM<br>Validation : NA | (Gräber et al., 2011) |
| - | • Linear static grey-box models<br>• Energy balance<br>• Constant heat recovery ratio and COP | CC | • 1 state | Application in an NLP based economic-MPC for optimal scheduling of primary HVAC equipment<br>Validation : NA | (Zhao et al., 2015) |
| **CHP:**<br><br>Gas engine | • Linear static grey-box models<br>• Linear interpolation or quadratic regression of predefined parameters for different load factors<br>• Energy balance | Cost of operation, power, fuel | • > 2 parameters | Application in an MILP or MINLP for cost based optimisation to design and operate a CCHP system in a simulation environment<br>Validation : NA | (Ren et al., 2008), (Cho et al., 2009a), (Bracco et al., 2013) |
| Gas turbine with internal PID control | • Nonlinear static black-box models<br>• Regression based static input-output relationships<br>• Reduced order modelling | Fuel, exhaust gas mass flow & temperature | • 3 parameter sets(10 total) | Application in an NLP to solve a cost based look-ahead optimisation problem for operating strategy<br>Validation : Visual | (Chandan et al., 2012) |
| Gas engine microCHP | • Nonlinear dynamic input-output black-box models<br>• Transfer function using step-response analysis | Power | • 1 state | Application in active voltage management using a proportional integral controller<br>Validation : Visual | (Hidalgo Rodriguez et al., 2012) |
| Gas engine microCHP with internal mass flow controller | • Nonlinear dynamic grey-box models<br>• Curve fits for part-load behaviour<br>• Mass and energy balance over engine and heat exchanger | CT, power, fuel, variable mass flow | • > 5 component parameters<br>• 3 parameter sets (fitting coefficients) | Simulation of the dynamic behaviour of a MicroCHP including its part-load behaviour and internal control logic<br>Validation : Visual | (Seifert, 2013) |
| **OC:**<br>Counter flow wet cooling tower | • Merkel's Theory<br>• NTU-effectiveness method<br>• Levenberg-Marquardt method for parameter identification | Heat rejection rate | • 3 parameters<br>• 1 state | Application in real time optimisation of cooling tower operation<br>Validation : Visual, quantitative (RMSRE) | (Jin et al., 2007) |
| Counter flow wet/dry cooling tower | • NTU-effectiveness method<br>• Mass and energy balance | Heat rejection rate | • 4 parameters | Application in simulation of cooling tower performance for design purposes | (Bergman et al., 2011; Mitchell and Braun, 2013) |
| **RevHP:**<br><br>Air cooled electric compressor chiller | • Nonlinear static grey-box models<br>• Parameter estimation from catalogue data<br>• Pressure-enthalpy based mass and energy balance over chiller internal components | CC, PI, CT and COP | • > 10 Parameters | Deploy in energy calculation and/or building simulation programs to simulate detailed behaviour (including control logic) of an electric chiller<br>Validation: Visual, quantitative (RMS error) | (Jin and Spitler, 2002), (Lemort et al., 2009) |
| - | • Nonlinear static grey-box models<br>• Look up tables of manufacturer data<br>• Energy balance | CC, PI, CT and COP | • > 5 parameter sets | Nonlinear optimisation of CCHP operation using mass flow and chilled water temperature as variables<br>Simulation of a GSHP<br>Validation: Visual, quantitative (MAPE) | (Ma et al., 2009), (Salvalai, 2012) |
| Electrical Chiller | • Nonlinear dynamic grey-box models<br>• Polynomial fit of COP to Carnot efficiency<br>• Mass and energy balance over chiller internal components | CC, PI, CT and COP | • > 10 Parameters<br>• 1 Curve fit | Building HVAC system simulation within a Modelica environment<br>Validation: NA | (Wetter, 2009) |
| Compression chiller (Electrical and mechanical) | • Nonlinear static black-box models<br>• Model identification using least squares | CC and CT | • > 5 parameters | Development of a controller for the chiller's variable speed compressor, DP for optimal scheduling of a CCHP in a simulation environment<br>Validation : Visual, quantitative (percentage error) | (Romero et al., 2011), (Facci et al., 2014) |
| Water cooled electric compressor chiller | • Nonlinear static grey-box models<br>• Pressure-enthalpy based thermodynamic balance<br>• Semi-empirical models of chiller internal components | CC, PI, CT and COP | • > 10 parameters | Developing a practical model based supervisory control of a HVAC plant comprising of this chiller.<br>Validation : Visual, quantitative (RMS error and MAE) | (Jin et al., 2011) |
| Electrical Chiller | • Linearised state-space dynamic models<br>• Mass and energy balance over chiller internal components | CC, PI, CT and COP | • 6 states<br>• > 20 parameters | Simulation of the chiller's transient behaviour under different perturbations and initial conditions<br>Validation : Quantitative (AE) | (Yao et al., 2013) |
| **HTES/CTES:**<br><br>Stratified | • 1-D dynamic multilayer model<br>• Fourier's equation for heat flow<br>• Mass and energy balance per layer<br>• If-else logic for charging/discharging<br>• Effective vertical heat conductivity | Temperature distribution | • 1 state per layer<br>• < 8 parameters | Simulation of the transient temperatures in a stratified tank<br>A simplified 2-layer model with 2 states per layer applied in nonlinear MPC for scheduling of a chiller plant in a simulation environment.<br>Validation : Visual | (Eicker, 2006), (Dwivedi, 2009), (Ma et al., 2009) |
| Mixed | • Mass and energy balance<br>• Figure of merit concept<br>• Tank charging and discharging rate as control variable | State-of-charge of tank and tank temperature | • 1 state | Application in an MILP algorithm for chilling plant design optimisation or in an MINLP algorithm for optimal design of a DSH network<br>Validation : NA | (Henze et al., 2008), (Tveit et al., 2009) |
| Stratified | • Multinode model with heat conduction only | Temperature distribution | • < 10 Parameters | Building HVAC system simulation within a Modelica environment<br>Validation : NA | (Wetter, 2009) |
| Stratified | • Nonlinear dynamic DAE model for a 2 layer tank<br>• Mass and energy balance per layer<br>• Regression based time delays | Temperature distribution | • > 10 Parameters<br>• 2 states | Nonlinear optimisation of CCHP operation<br>Validation : Visual | (Chandan et al., 2012) |

AE (Absolute error), APE (Absolute percentage error), CC (Cooling capacity), CCHP (Combined cooling heating and power), COP (Coefficient of performance), CT (Circuit temperatures), DAE (Differential Algebraic Equations), DP (Dynamic programming), DSH (District heating), GSHP (Ground source heat pump), LDF (Linear driving force), MAE (Mean absolute error), MAPE (Mean absolute percentage error), MILP (Mixed integer linear program), MINLP (Mixed integer nonlinear program), NA (Not available), NLP (Nonlinear program), NSD (Normalised standard deviation), NTU (Number of Transfer Units), PI (Power Input), RMS (Root mean squared), RMSRE (Root mean squared relative error)

Table 3 Qualitative analysis of selective studies.

| | Reference (Years Ascending) | Component Dynamics | Part-Load Behaviour or Internal Control Logic | Live Parameter Identification Capabilities | Accuracy | Continuous Differentiability | Generalisation Capabilities | Adaptability to Component Design | Complexity |
|---|---|---|---|---|---|---|---|---|---|
| AdCM | (Li and Wu, 2009) | Yes | Yes / Yes | No | Very high | Yes | Medium | - | Very high |
| | (Schicktanz and Núñez, 2009) | Yes | Yes / Yes | No | Very high | Yes | Medium | - | Very high |
| | (Gräber et al., 2011) | Yes | Yes / Yes | No | Very high | Yes | Medium | - | Very high |
| | (Zhao et al., 2015) | No | No / No | No | Very low | Yes | Low | - | Very low |
| CHP | (Ren et al., 2008) | No | Yes / No | No | High | Yes | Medium | - | Low |
| | (Hidalgo Rodriguez et al., 2012) | Yes | No / No | Yes | Very high | Yes | Low | - | Low |
| | (Bracco et al., 2013) | No | Yes / No | No | High | Yes | Medium | - | Low |
| | (Seifert, 2013) | Yes | Yes / Yes | Yes | High | Yes | High | - | High |
| | (Li et al., 2014) | No | Yes / No | No | High | Yes | Medium | - | Low |
| RevHP | (Jin and Spitler, 2002) | No | Yes / - | No | High | Yes | Medium | - | High |
| | (Ma et al., 2009) | No | Yes / - | No | High | Yes | Medium | - | Medium |
| | (Wetter, 2009) | Yes | Yes / - | No | Very high | Yes | High | - | High |
| | (Salvalai, 2012) | No | Yes / - | Yes | High | Yes | High | - | Medium |
| | (Facci et al., 2014) | No | Yes / - | No | Very high | Yes | Low | - | Low |
| HTES | (Eicker, 2006) | Yes | - | - | High | No | Medium | No | Medium |
| | (Henze et al., 2008) | Yes | - | - | Low | No | High | No | Low |
| | (Ma et al., 2009) | Yes | - | - | Low | No | High | No | Low |
| | (Wetter, 2009) | Yes | - | - | Very high | No | High | Yes | High |
| | (Chandan et al., 2012) | Yes | - | - | Very high | Yes | Medium | No | High |

In the grey-box modelling approach, we used the regression analysis and the step-response analysis methods for fitting data and determining the dynamic properties of the components. Both these methods make it possible for the user to choose data sets that are either readily available in the manufacturer's catalogues or can be collected during commissioning of the equipment.
The basic theory of regression analysis and step-response analysis is shown below:

### *3.1 Regression Analysis*

Regression analysis is a method to find a polynomial relationship among response or dependent variables and explanatory or independent variables. A regression analysis is linear when the polynomial is linear in the coefficients. However, the regression could be univariate if only one independent variable exists as in (1) or multivariate if multiple independent variables are considered as in (2). The polynomial itself could be a first order or a higher order polynomial depending on the characteristics of the data that is being fit (Fumo and Rafe Biswas, 2015).

$$y^* = \beta_0 + \beta_1 x_1 + \beta_2 {x_1}^2 \tag{1}$$

$$y^* = \beta_0 + \beta_1 x_1 + \beta_2 x_2 + \beta_3 x_1 x_2 + \beta_4 {x_1}^2 + \beta_5 {x_2}^2 \tag{2}$$

where,
$y^*$ – Dependent variable (predicted value in models)
$\beta_1, \beta_2 \ldots \beta_5$ – Coefficients of regression
$x_1, x_2$ – Independent variables
Using (1) and (2), a polynomial regression analysis could be carried out to minimise a normalised sum of squared error problem as shown in (3) for fitting apriori data and estimating the coefficients of regression.

$$min\, J = \sum_{i=1}^{n} \left( \frac{y_i - y_i^*}{y_i} \right)^2 \tag{3}$$

where,
$J$ – Cost function
$y_i$ – i[th] measured value
$y_i^*$ – i[th] predicted value
The application of the method is described in the respective part of the individual component model.

### *3.2 Step-Response Analysis*

The dynamic response of a controlled system can be described via the manipulated variable step-response or the interference variable step-response. A manipulated variable step-response is more common in practice and is characterised by the time constant $T_S$ and the transfer coefficient or gain $K_S$. In building technologies, very often the behaviour of a first order system or system with one storage element (PT-1 element) is observed. The input/output differential equation for this system is given in (4).

$$\dot{y}_i^* + \frac{1}{T_S} y_i^* = \frac{k_S}{T_S} u \tag{4}$$

where,
$y_i^*$ – System output (predicted value in models)
$u$ – Manipulated variable ($Switch_{Component}$)
$K_S$ – Transfer coefficient
$T_S$ – Time constant

The application of the method is described in the respective part of the individual component model.
The simulation models are developed in OpenModelica as input / output equations for the individual components using elemental libraries like SI Units and connectors as shown in *Fig. 3* (Fritzson, 2014).

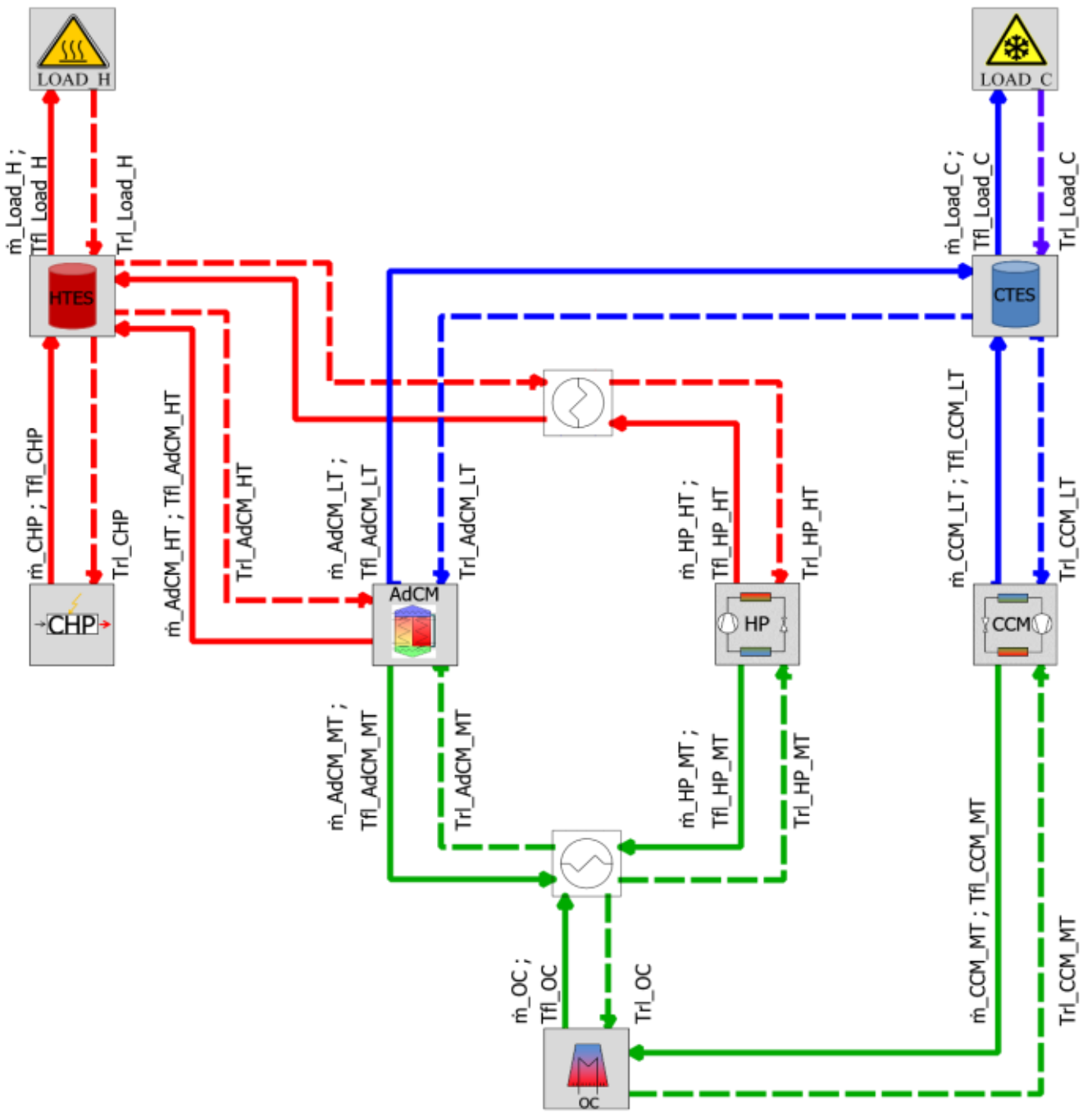

Fig. 3 Model of the INES polygeneration system in OpenModelica.

Nomenclature from process engineering is followed, wherein flows leaving a component are designated *feed-line* (subscript "fl"), and flows entering a component are designated *return-line* (subscript "rl"). In addition, based on circuit-temperature, *high temperature circuit* is denoted with a subscript "_HT", *medium temperature circuit* with "_MT", and *low temperature circuit* with "_LT".
For the regression analysis, the generalised reducing gradient search (via Microsoft Excel's Data Solver®) algorithm was applied and for the step-response analysis, the "Control Design and Simulation Module" in LabVIEW® was used.

## 4. MODELS OF THE PRIMARY SYSTEM COMPONENTS

For the modelling of this complex energy plant, simplifying assumptions based on engineering intuition and apriori knowledge of the system were made for reducing the computing time of the target optimisation problem. Component specific assumptions are described in their relevant segment and the general assumptions are listed below:

- Heat losses and pressure losses through pipes and components (other than storages) were neglected,
- specific heats and densities of all fluids were assumed constant,
- ideal conservation of mass was assumed,
- at full load real power is the nominal power of the equipment,
- internal controllers of components are ideal and reliable,

- volume flows in the circuits are constant (other than for CHP and loads) and maintained at nominal flows,
- accurate forecast of ambient temperature and building thermal loads are available over the entire simulation horizon.

The current mass flow in the hydraulic circuits, induced by a machine is computed depending on the machine's operation status as follows (5):

$$\dot{m}_{Component} = Switch_{Component} \frac{\dot{v}_{Component}\rho_{fluid}}{3600} \quad (5)$$

This formulation achieves two results; firstly, when the machines are switched off, then no mass flow occurs between components and storages ensuring that their temperatures are not affected. Secondly, the switches of the components, which will be the decision variables in the optimisation problem, occur a reduced number of times in the MPC models since they need not be used in all the mass and energy balance equations for the individual models. Thereby, avoiding redundancies of the decision variables in the optimisation problem.

In the following sub-sections, the individual models that were developed applying the grey-box modelling methodology are described.

### *4.1 Adsorption Chilling Machine (AdCM) model*

An AdCM operates on the principle of sorption of solids (adsorption) like silica-gel and the cooling effect is produced by an adsorption triggered evaporation (Chua et al., 1999). This is the cooling capacity of the machine denoted by $P_{th_AdCM_LT}$ in the Low Temperature (LT) circuit. For the desorption process, the driving heat could be provided by sources like solar thermal, CHPs or industrial process heat and is denoted by $P_{th_AdCM_HT}$ in the High Temperature (HT) circuit. Lastly, the heat created during adsorption and liquefaction is discharged to the environment over the outdoor coil and is denoted as $P_{th_AdCM_MT}$ in the Medium Temperature (MT) circuit. For continuous cooling, modern aggregates consist of two process modules that are interconnected through automatic 3-way-valves. Each module goes through the adsorption/desorption and heat and mass recovery phases. This complex operation is achieved through internally controlled switching of valves and pumps using complex control algorithms leading to the distinctive cyclic behaviour of the three circuit temperatures in an AdCM, as reported in experimental set-ups in literature (Núñez, 2010). The modelling of this internal dynamics is extremely complex and has been included in some models as shown in *Table 2.* Then again, the complexity of these models make them ineffective for a system wide optimisation and simplified models of an AdCM should be developed. Another extreme is a highly simplified linear energy balance model which assumes a constant Coefficient of Performance (COP) and does not capture the part-load behaviour of the machine even though it is extremely important because of the sensitive dependence of the AdCM's performance on its inlet temperatures. As seen in Section 2, a balance of complexity and accuracy must be achieved to develop a practical AdCM model that is part of an entire system being optimally scheduled.

Based on our previous experimental work and literature research, we established important simplifying assumptions for the modelling of this component:

- for typical AdCM based energy systems, there are always adequate storages (hot and cold) planned and these smoothen the cyclic temperature pattern due to their damping effect, thus making it unnecessary to model this pattern in detail (Bürger et al., 2017; Sawant and Pfafferott, 2017)
- the heat released to the environment over an entire cycle is approximately equal to the sum of driving heat and the cooling capacity (Bürger et al., 2017; Sawant and Pfafferott, 2017)
- manufacturer's catalogues of widely used industrial AdCMs provide characteristic curves for cooling capacity and COP based on inlet temperatures in the three circuits (FAHRENHEIT GmbH, 2019; Invensor GmbH, 2019)

Considering the above findings and assumptions, we applied regression analysis to fit the cooling capacity and $COP$ of the AdCM as second order functions of the three inlet temperatures as shown in (6) and (7).

$$P_{th_AdCM_LT} = d_1 + d_2.T_{rl_AdCM_LT} + d_3.T_{rl_AdCM_HT} + d_4.T_{rl_AdCM_MT} + d_5.T_{rl_AdCM_LT}^2 + d_6.T_{rl_AdCM_HT}^2 + d_7.T_{rl_AdCM_MT}^2 + d_8.T_{rl_AdCM_LT}T_{rl_AdCM_HT} + d_9.T_{rl_AdCM_LT}T_{rl_AdCM_MT} + d_{10}.T_{rl_AdCM_HT}T_{rl_AdCM_MT} \quad (6)$$

$$COP = e_1 + e_2.T_{rl_AdCM_LT} + e_3.T_{rl_AdCM_HT} + e_4.T_{rl_AdCM_MT} + e_5.T_{rl_AdCM_LT}^2 + e_6.T_{rl_AdCM_HT}^2 + e_7.T_{rl_AdCM_MT}^2 + e_8.T_{rl_AdCM_LT}T_{rl_AdCM_HT} + e_9.T_{rl_AdCM_LT}T_{rl_AdCM_MT} + e_{10}.T_{rl_AdCM_HT}T_{rl_AdCM_MT} \quad (7)$$

The energy balance over the three circuits was done as per (8) and (9).

$$P_{th_AdCM_HT} = \frac{P_{th_AdCM_LT}}{COP} \quad (8)$$

$$P_{th_AdCM_MT} = P_{th_AdCM_HT} + P_{th_AdCM_LT} \quad (9)$$

Using the calculated thermal powers and applying the first law of thermodynamics, the feed-line temperatures for each circuit were calculated as in (10), (11) and (12).

$$T_{fl_\mathrm{AdCM_LT}} = T_{rl_\mathrm{AdCM_LT}} - \frac{P_{th_\mathrm{AdCM_LT}}}{\frac{\rho_w}{3600}\dot{v}_{AdCM_LT}c_{p_w}} \quad (10)$$

$$T_{fl_\mathrm{AdCM_MT}} = T_{rl_\mathrm{AdCM_MT}} + \frac{P_{th_\mathrm{AdCM_MT}}}{\frac{\rho_w}{3600}\dot{v}_{AdCM_MT}c_{p_w}} \quad (11)$$

$$T_{fl_\mathrm{AdCM_HT}} = T_{rl_\mathrm{AdCM_HT}} - \frac{P_{th_\mathrm{AdCM_HT}}}{\frac{\rho_w}{3600}\dot{v}_{AdCM_LT}c_{p_w}} \quad (12)$$

By applying the volume flow term $\dot{v}_{AdCM}$ in the equations above instead of the mass flow, a numerical error in simulations is avoided such that a division by zero does not occur when the machines are turned off and mass flows are zero. The mass flows in the three circuits are calculated using (5) shown earlier.

Finally, the electric consumption of the AdCM is given by (13).

$$P_{el_AdCM} = Switch_{AdCM}P_{\mathrm{el}_AdCM_\mathrm{Nom}} \quad (13)$$

At the end of this section, the coefficients of regression are given in *Table 4* and the Information Flow Diagram (IFD) for the AdCM model is shown.

*4.2 Combined Heating and Power (CHP) model*

The CHP comprises of a single cylinder engine coupled to an asynchronous generator that convert fuel into thermal $P_{th_CHP}$ and electrical power $P_{el_CHP}$. The heat is transferred to the cooling water of the CHP, which flows in a closed circuit connected to the stratified Hot Thermal Energy Storage Tank (HTES). Colder water coming from the bottom of the HTES enters in the return-line of the CHP with $T_{rl_CHP}$ and hotter water leaving the CHP at $T_{fl_CHP}$ enters in the feed-line and is added to the top of the HTES. An integrated controller in the CHP maintains the following internal control logic:

- volume flow of water $\dot{v}_{CHP}$ is controlled depending on return line temperature $T_{rl_CHP}$ to minimise part-load losses
- start-up checks needing 25 s (introduce delay time of approx. 25 s).

Further analysis of the functional tests of the CHP showed slow dynamics for the $P_{th_CHP}$ similar to a PT-1 element during start-up (Sawant and Pfafferott, 2017).

As shown in *Table 3*, most models used in literature for optimisation do not integrate this control logic or dynamic behaviour and are typically linear fits of apriori data. Some models use the black-box approach requiring many high quality data sets for parameterisation and thus making it difficult to generalise the models for other systems. Certain models depicting the dynamic behaviour take the approach of a mass and energy balance over the engine block and the heat exchanger thereby increasing the number of system-states and parameters for modelling the CHP. This introduces greater complexity in the optimisation problem.

Based on our previous experimental work and literature research, we have established important simplifying assumptions for the modelling of this component:

- the delay time after start-up can be neglected since the length of the sampling time and forecast horizon for a 15-minute electricity price based MPC is significantly larger than the delay time interval itself,
- the internal control logic of the varying $\dot{v}_{CHP}$ can be portrayed using the regression based approach where the $\dot{v}_{CHP}$ is fit to the incoming $T_{rl_CHP}$ using a second order univariate linear regression (Seifert, 2013),
- the dynamic behaviour of the thermal power $P_{th_CHP}$ can be portrayed using a differential equation obtained by the step-response analysis method shown in Section 3.2 (Hidalgo Rodriguez et al., 2012),
- internal control logic of the modern day CHPs and their operation in combination with storages and other components ensure close to full-load operation and thus, if this logic is included in the model, it is not necessary to simulate the part-load operation separately and constant efficiencies can be assumed for optimisation problems (Zhou et al., 2013),
- Higher Calorific Value (HCV) of fuel is used for calculation,
- a complete combustion of fuel occurs in the CHP.

Considering the above findings and assumptions, we modelled the important characteristics of the CHP's operation as shown in (14) and (15).

$$\dot{v}_{CHP} = b_1 + b_2.T_{rl_CHP} + b_3.T_{rl_CHP}^2 \quad (14)$$

$$\dot{P}_{th_CHP} = \frac{P_{th_CHP_Nom} Switch_{CHP}}{c_1} - \frac{P_{th_CHP}}{c_1} \quad (15)$$

Here $c_1$ represents the average time constant (in seconds) of the CHP system that was determined by performing a $Switch_{CHP}$ step-response analysis over three tests with varying initial temperatures and is given in *Table 4*.

Using the calculated thermal power and volume flow and applying the first law of thermodynamics the feed-line temperature was calculated as in (16)

$$T_{fl_CHP} = T_{rl_CHP} + \frac{P_{th_CHP}}{\frac{\rho_W}{3600}\dot{v}_{CHP} c_{p_w}} \quad (16)$$

The mass flow going to the HTES was calculated using (5) shown earlier and the electrical production of the CHP $P_{el_CHP}$ is given below:

$$P_{el_CHP} = Switch_{CHP} P_{el_CHP_Nom} \quad (17)$$

Furthermore, the fuel consumed by the CHP was calculated using (18). This formulation aids in generalising the type of fuel that could be used in the simulation.

$$\dot{v}_{Fuel} = Switch_{CHP} \frac{(P_{el_CHP_Nom} + P_{th_CHP_Nom})}{HCV_{Fuel} \cdot (\eta_{el_Nom} + \eta_{th_Nom})} \quad (18)$$

At the end of this section, the coefficients of regression are given in *Table 4* and the IFD for the CHP model is shown.

*4.3 Outdoor Coil (OC) model and heat exchangers*

As seen in *Fig. 3* the outdoor coil is the heat-sink for the condenser of the chilling machines (AdCM and CCM) and the heat-source for the evaporator of the Heat Pump (HP). It is principally a dry-cooling tower with three variable-speed fan motors consuming a total $P_{el_OC_max}$ of 0.9 $kW_{el}$ at their maximum speed, $RPM_{max}$ of 480 RPM. The actual speed of the fans $RPM$ can be controlled with a 0 – 10 volt signal $V_{set_OC}$. The total heat exchanger area $A$ is 521.8 m$^2$. The fluid in the circuit is a 34% glycol-water mixture (brine).

The OC and heat exchanger models are motivated from the "Number of Transfer Units – Effectiveness (NTU-ε)" method (Bergman et al., 2011). Additionally, we applied the "Fan Affinity Laws" to establish the relationship between the $RPM$, the mass flow of air $\dot{m}_{air}$ and electrical power consumed by the OC $P_{el_OC}$ as seen in (20) and (21) (Wagner and Gilman, 2011). Here we assumed a directly proportional behaviour of fan speed with respect to the volt signal as seen in (19)

$$RPM = \frac{RPM_{max} V_{set_OC}}{V_{set_OC_max}} \quad (19)$$

$$\dot{m}_{air} = \frac{RPM \dot{m}_{air_max}}{RPM_{max}} \quad (20)$$

$$P_{el_OC} = Switch_{OC} \frac{RPM^3 P_{el_OC_max}}{RPM_{max}^3} \quad (21)$$

Other assumptions for this model are:

- homogeneous air flow,
- effect of the instantaneous variations of air speed on the pressure is neglected,
- no pressure loss over the heat exchangers.

In this paper, the application of this method for simulating the OC and the heat exchangers is explained with the help of the OC model only. The NTU-ε method calculates the effectiveness of a heat exchanger based on the maximum possible heat transfer that can be hypothetically achieved. The heat capacity rates for the hot (brine) and cold (air) fluids are denoted as $C_h$ and $C_c$ respectively and calculated as shown in equations (22) and (23).

$$C_h = \dot{m}_{OC} c_{p_\mathrm{b}} \tag{22}$$

$$C_c = \dot{m}_{air} c_{p_air} \tag{23}$$

Where, $c_{p_air}$ and $c_{p_\mathrm{b}}$ are the specific heat capacities of the fluids and $\dot{m}_{air}$ and $\dot{m}_{OC}$are their mass flows.

$C_{min}$ ($C_{max}$) used later in this method is the smaller (larger) out of the two heat capacity rates. The maximum possible heat transfer, $P_{th_OC_max}$ for the OC per unit time is given by (24).

$$P_{th_OC_max} = C_{min}(T_{rl_OC} - T_{Amb}) \tag{24}$$

Applying (25), (26) and (27) the effectiveness of a cross-flow heat exchanger ε was calculated as follows:

$$NTU = \frac{UA}{C_{min}} \tag{25}$$

$$C_r = \frac{C_{min}}{C_{max}} \tag{26}$$

$$\varepsilon = \frac{1-\mathrm{e}^{[-NTU(1-C_r)]}}{1-C_r e^{[-NTU(1-C_r)]}} \tag{27}$$

As shown in (28), $\varepsilon$ is applied to calculate the thermal power $P_{th_OC}$ of the OC assuming it to be an air-fluid heat exchanger

$$P_{th_OC} = \varepsilon\, P_{th_OC_max} \tag{28}$$

By means of the above equations and with energy balance over the OC, we get the two outlet temperatures from (29) and (30).

$$T_{fl_\mathrm{OC}} = T_{\mathrm{rl_OC}} - \frac{P_{th_OC}}{C_h} \tag{29}$$

$$T_{air_\mathrm{out}} = T_{Amb} + \frac{P_{th_OC}}{C_c} \tag{30}$$

As mentioned before, the mass flow from the OC to the other components was calculated using (5). At the end of this section, the coefficients of regression are given in *Table 4* and the IFD for the OC model is shown.

### *4.4 Reversible Heat Pump (RevHP) model*

The RevHP can operate as a Heat Pump (HP) or as a Compression Chilling Machine (CCM) and is principally a conventional refrigeration system consisting of the four key components (i.e. compressor, condenser, expansion valve and evaporator) operating on the vapour-compression cycle (Salvalai, 2012; Yao et al., 2013). As seen in *Fig. 3*, the heating or cooling effect is generated when:

- the OC acts as a heat-source ($P_{th_\mathrm{HP_MT}}$) to evaporate the refrigerant (heating operation) and the compressor input ($P_{el_\mathrm{RevHP}}$) increases its temperature and pressure which is eventually released as the heating capacity ($P_{th_\mathrm{HP_HT}}$),
- the OC acts as a heat-sink ($P_{th_\mathrm{CCM_MT}}$) to condense the high temperature and high pressure compressed vapour (cooling operation) which was evaporated using the heat from the loads leading to a cooling effect ($P_{th_\mathrm{CCM_LT}}$).

The switching in the operation is done either internally with a reversing valve or over the external hydraulic connections from the evaporator/condenser to the OC.

As shown in *Table 3*, we found modelling approaches in the literature that use data tables from manufacturers or data that is readily available during the commissioning of these machines.

Using such empirical data for our machine (Daikin Europe, 2016) a second order equation like (31), (32) and (33) was fit using polynomial regression for calculating $P_{th_\mathrm{HP_HT}}$, $P_{th_\mathrm{CCM_LT}}$ and $P_{el_\mathrm{RevHP}}$ as a function of the inlet temperatures in the condenser (subscript "_c") and evaporator (subscript "_e") circuits.

$$P_{th_HP_HT} = g_1 + g_2 T_{rl_HP_HT} + g_3 T_{rl_HP_MT} + g_4 T_{rl_HP_HT} T_{rl_HP_MT} + g_5 T^2_{rl_HP_HT} + g_6 T^2_{rl_HP_MT} \tag{31}$$

$$P_{th_CCM_LT} = h_1 + h_2 T_{rl_CCM_MT} + h_3 T_{rl_CCM_LT} + h_4 T_{rl_CCM_MT} T_{rl_CCM_LT} + h_5 T^2_{rl_CCM_MT} + h_6 T^2_{rl_CCM_LT} \tag{32}$$

$$P_{el_RevHP} = Switch_{Rev_HP}(i_1 + i_2 T_{rl_RevHP_e} + i_3 T_{rl_RevHP_c} + i_4 T_{rl_RevHP_e} T_{rl_RevHP_c} + i_5 T^2_{rl_RevHP_e} + i_6 T^2_{rl_RevHP_c}) \tag{33}$$

Assuming an ideal refrigeration cycle, the energy balance for the RevHP in the HP mode was calculated by (34) and in the CCM mode by (35) (Sawant and Doan, 2017). Equation (36) gives the $COP$ for the RevHP in general:

$$P_{th_HP_MT} = P_{th_HP_HT} - P_{el_RevHP} \tag{34}$$

$$P_{th_CCM_MT} = P_{th_CCM_LT} + P_{el_RevHP} \tag{35}$$

$$COP = \frac{P_{th_RevHP}}{P_{el_RevHP}} \tag{36}$$

The first law of thermodynamics was applied in each circuit to get the feed-line temperatures as shown below for the HP condenser circuit and the CCM evaporator circuit:

$$T_{fl_HP_HT} = T_{rl_HP_HT} + \frac{P_{th_HP_HT}}{\frac{\rho_W}{3600}\dot{v}_{HP_HT} c_{p_b}} \tag{37}$$

$$T_{fl_\mathrm{CCM}_LT} = T_{rl_\mathrm{CCM}_LT} - \frac{P_{th_\mathrm{CCM}_LT}}{\frac{\rho_W}{3600}\dot{v}_{CCM_LT} c_{p_w}} \tag{38}$$

The mass flows in the circuits were calculated using the formulation in (5) discussed earlier and the coefficients of regression and IFDs are shown at the end of this section.

### *4.5 Thermal Energy Storage (HTES & CTES) model*

In simple words, thermal storages help to balance out the mismatch in the production and consumption cycles especially in variable renewable energy systems. However, the modelling of such storages is complex due to physical effects of thermal stratification, forced convection or laminar flows that may occur based on the construction of the tank. The simulation of the stratification effect is important as this increases the performance of the tank and it is closely linked with the dynamic operation of the plant especially when performing cost based operational optimisation (Campos Celador et al., 2011). In the literature (*Table 2 & Table 3*), most works focusing on optimisation for design of energy systems apply mixed storage tanks, however they highly recommend to continue further research for optimal scheduling applying at least a simple stratified tank model (Campos Celador et al., 2011; De Césaro Oliveski et al., 2003)

The model of the thermal storages in this work is based on a 1-D dynamic multilayer model using the Fourier's equation (Eicker, 2006; Streckiene et al., 2011). This analytical model summarises the complex thermal flux using an effective vertical heat conductivity coefficient $\lambda_{eff}$. The HTES is considered as a vertically stratified cylindrical tank as shown in *Fig. 4* with user defined dimensional parameters such as the diameter D, height H, thickness of tank wall Th and number of layers in the longitudinal direction N. An effective mass flow $\dot{m}_i$ for each layer is calculated based on the balance of mass flows from the source circuit (subscript "*s*") and load circuit (subscript "*l*"). A positive $\dot{m}_i$ with energy entry from layer above the i$^{th}$ layer (i+1) is interpreted by the binary parameter $\delta_i^+$ = 1, otherwise $\delta_i^+$ = 0. A negative $\dot{m}_i$ from layer below the i$^{th}$ layer (i-1), i.e. dominance of the load mass flow and thus cooling of layer i, is taken into account by the parameter $\delta_i^-$. Only for the top most layer (N$^{th}$ layer) with hot water entering from the feed-line of the source circuit, the parameter $\delta_i^s$= 1 and analogously for the bottom most layer (1$^{st}$ layer) with cooler water entering from the return-line of the load circuit the parameter $\delta_i^l$= 1. From the user defined dimensional parameters of the tank other relevant dimensional quantities such as the exterior heat transfer surface area of a layer $A_{ext.i}$, cross-section area of a layer $A_i$, mass of a layer $m_i$ and height of a layer $z_i$ were calculated as follows:

$$z_i = H/N \tag{39}$$

$$A_{ext,i} = \pi D z_i \tag{40}$$

$$A_i = \pi\,(D - 2Th)^2/4 \tag{41}$$

$$m_\text{i} = A_i z_i \rho_w \tag{42}$$

The general energy balance of each layer is then calculated as shown in (43).

$$m_i c_p \dot{T} = \delta_i^s (\dot{m} c_p)_s (T_{fl_s} - T_i) - \delta_i^l (\dot{m} c_p)_l (T_i - T_{rl_l}) - kA_{ext,i}(T_i - T_{amb}) + \delta_i^+ \dot{m}_i c_p (T_{i+1} - T_i) + \delta_i^- \dot{m}_{i+1} c_p (T_i - T_{i-1}) + \frac{A_i \lambda_{eff}}{z_i}(T_{i+1} - 2T_i + T_{i-1}) \tag{43}$$

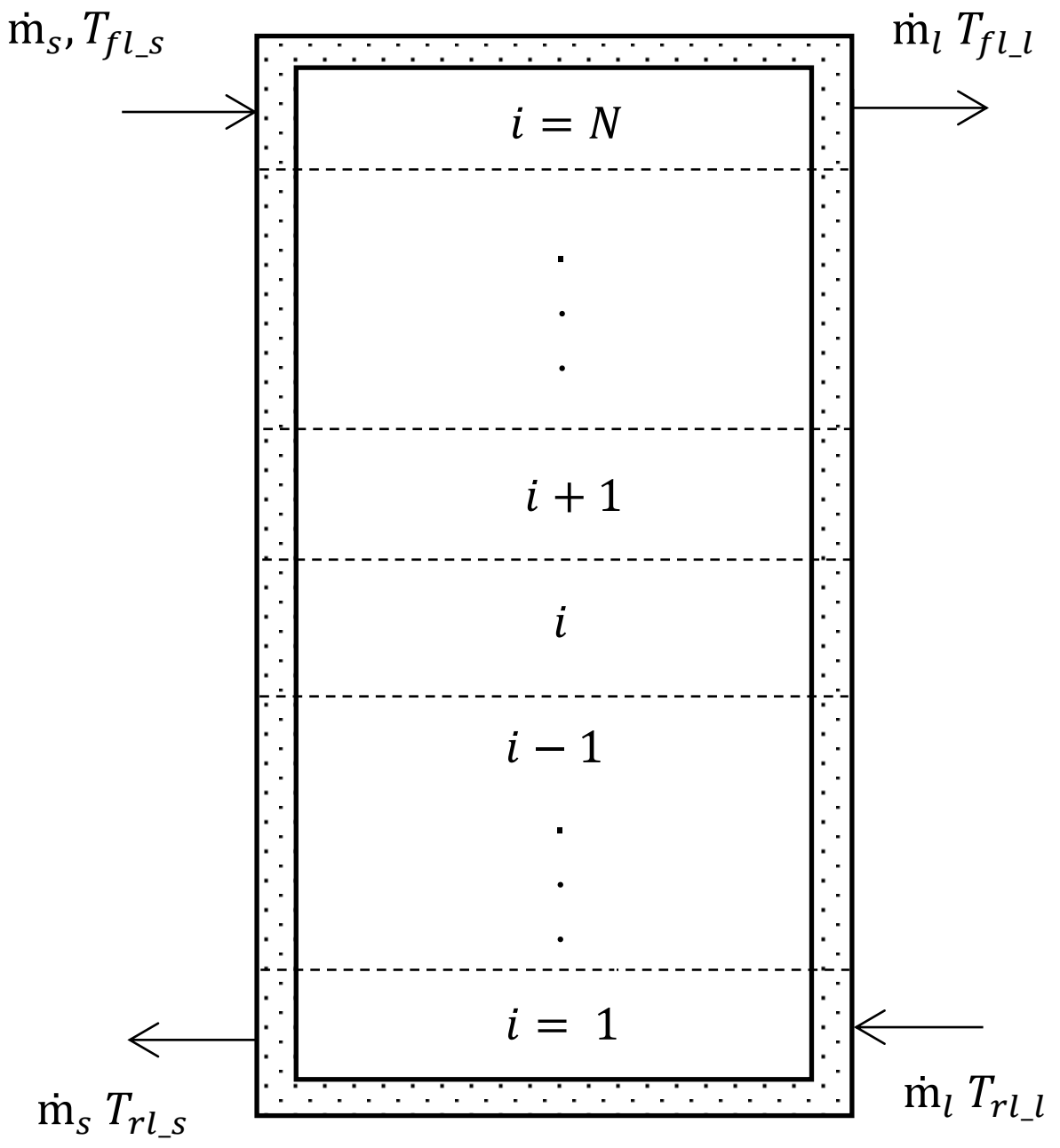

Fig. 4 Schematic depiction of the HTES with hydraulic connections and layer numbering.

where,
t - time (s)
$T$ - temperature of layer (°C)
$k$ - overall heat transfer coefficient (W/(m$^2$·K))
$\lambda_{eff}$ - effective vertical heat conductivity of water (W/(m.K))

For a well insulated steel tank $k$ and $\lambda_{eff}$ are assumed to be 0.002 W/(m$^2$·K) and 0.0015 W/(m.K) respectively (Eicker, 2006).

With a given initial temperature distribution, the differential equation is applied to each layer and integrated over the entire forecast horizon to calculate the analytical temperature distribution over that time period. The limitations of this approach for application in our scenarios are as follows:

- *Differentiability*: For application within gradient-based optimization methods, models must be continuous and differentiable (Bürger et al., 2018). The presence of "If-Else"-statements within models, however, introduces discontinuities and must therefore be avoided.

Therefore, we modified the formulation of the energy balance for each layer to avoid the "If-Else" condition based on the effective mass flows. Consider the following formulation:

$$\frac{p(q + r)}{2} + \frac{\left(\sqrt{p^2 + \omega}\right)(q - r)}{2}$$

Where,
$\text{p}, \text{q}, \text{r}, \omega \in \mathbb{R}$
and $\omega \ll p$.
For $p$ > 0 the formulation will take the value:
$\approx pq$
For $p$ < 0 the formulation will take the value:
$\approx -pr$

Drawing an analogy from this formulation and applying to the energy balance equation of each layer we developed (44)

$$m_i c_p \dot{T} = \delta_i^s (\dot{m} c_p)_s (T_{fl_s} - T_i) - \delta_i^l (\dot{m} c_p)_l (T_i - T_{rl_l}) - kA_{ext,i}(T_i - T_{amb}) + \frac{\dot{m}_i c_p (a + в)}{2} + \frac{\left(\sqrt{\dot{m}_i^2 + \omega}\right) c_p (a - в)}{2} + \frac{A_i \lambda_{eff}}{z_i}(T_{i+1} - 2T_i + T_{i-1}) \quad (44)$$

where,

$a = T_{i+1} - T_i$

$в = T_i - T_{i-1}$

The value for $\varepsilon$ should be far less than the magnitude of $\dot{m}_i$ when the source and load circuits are active. In this study, the $\dot{m}_i$ was in the range of 0.02 kg/s and 0.69 kg/s and a value of $2 \times 10^{-4}$ is presumed for $\omega$.

- *Component design*: For simplification purposes, it is assumed that the hot source water enters at the top of the tank and is delivered to the load from the top of the tank. Similarly, the bottom of the tank is connected to the source and load circuits. In reality the construction of a storage tank may have hydraulic connections at different heights of the tank and the user must have the possibility to define the respective layers of entry or exit of water. This will greatly improve the capability of the model to simulate the temperature distribution accurately and adapt the model to different constructions (Sawant et al., 2018).

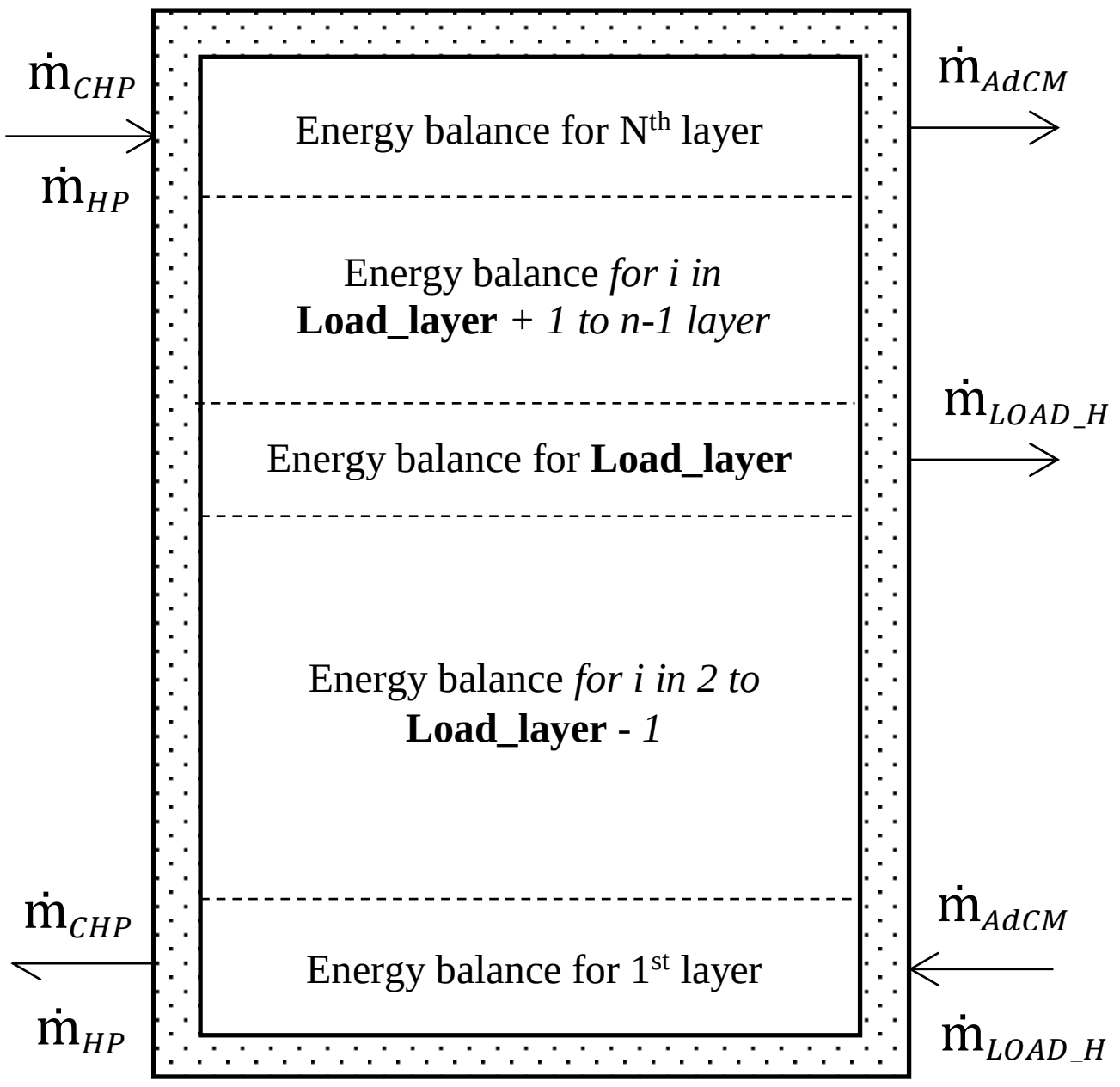

Fig. 5 Modification of tank model based on numerical loops to include a user defined parameter "Load_layer".

Here, we modified the formulation of the energy balance over the length of the tank by introducing a user-defined parameter "Load_layer". This acts as a tank splitting parameter and represents the layer from which water goes to the thermal loads. A differential equation is created for each section of the tank by implementing "For-loops" as shown in *Fig. 5*. Thus, by using an additional parameter a particular hydraulic connection was included in the model. This technique could be extended to multiple hydraulic connections at different heights of the tank.

The model of the CTES was similarly developed but adapted to the reversal of flows between the source and load circuits.

The tank models are discretised into 10 layers for each temperature sensor. Thus, the HTES with 9 temperatures sensors is discretised into 90 layers and the CTES with 4 sensors is discretised into 40 layers.

### *4.6 Thermal Loads (Load_H & Load_C) model*

Since the thermal loads are perfectly forecasted and are generated using the H-i-L set-up with a mixing valve, the models for the loads are developed by applying the first law of thermodynamics and the law of fluid mixing (Engineering ToolBox, 2011).

Under following assumptions:

- the return water temperature is same as the HVAC distribution element temperature assuming a high thermal conduction between the supply water and the distribution element,
- the feed line temperature $T_{fl_\mathrm{HVAC}}$ and mass flow $\dot{m}_{HVAC}$ in the HVAC circuit is constant and maintained by a field level three-point controller of a three-way mixing valve, the mass of water taken from the HTES for covering the heating load $P_{th_Load_H}$ was calculated using (45):

$$\dot{m}_{Load_H} = \frac{P_{th_Load_H} \dot{m}_{HVAC}}{\dot{m}_{HVAC} c_{p_w} (T_{fl_Load_H} - T_{fl_HVAC}) + P_{th_Load_H}} \quad (45)$$

Similarly, the mass flow from CTES $\dot{m}_{Load_C}$ to cooling load $P_{th_Load_C}$ was calculated based on an energy balance in the cooling circuit.

### *4.7 Data fitting coefficients*

The following table summarises all the fitting coefficients used in the modelling of the components.

Table 4

| Model | Coefficients | | | | | | | | | | |
|---|---|---|---|---|---|---|---|---|---|---|---|
| | | 1 | 2 | 3 | 4 | 5 | 6 | 7 | 8 | 9 | 10 |
| AdCM | d | 3.66 | 0.49 | 0.252 | -0.6 | 0.003 | 0.0 | 0.014 | 0.01 | -0.03 | -0.004 |
| | e | 0.42 | -0.02 | 0.006 | 0.002 | -0.001 | 0.0 | -0.001 | 0.0 | 0.002 | 0.0 |
| CHP | b | 0.43 | -0.15 | 0.0002 | | | | | | | |
| | c | 560.78 | | | | | | | | | |
| RevHP | g | 9.0 | 0.06 | 0.29 | 0.002 | -0.001 | -0.001 | | | | |
| | h | 9.0 | 0.04 | 0.30 | 0.002 | -0.002 | -0.001 | | | | |
| | i | 1.83 | -0.007 | 0.019 | 0.0 | 0.0 | 0.0 | | | | |

### *4.8 Information Flow Diagram (IFD)*

The inputs, outputs, and parameters of the models are listed in their IFDs in *Fig. 6*.

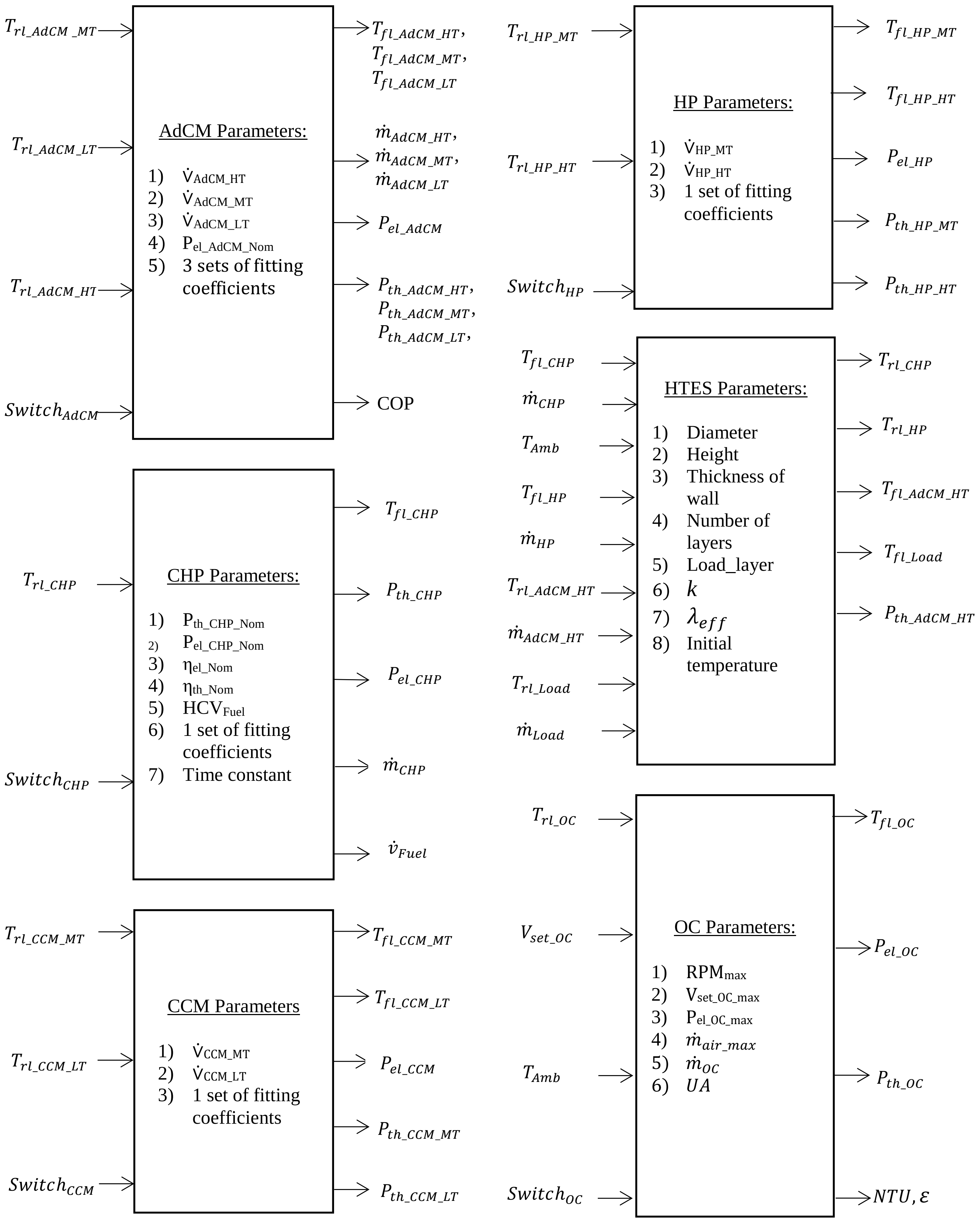

Fig. 6 Information flow diagrams of the models.

## 5. MODEL EVALUATION RESULTS

In the scope of this work, the plausibility of the models with respect to implementation in an optimal scheduling problem was evaluated using the process of empirical tests, where experimental results and simulation results were analysed visually and quantitatively. Experimental data from functional tests in the lab for four main operational modes under varying operational conditions (ambient temperatures, initial tank temperatures, load profiles etc.) were used for the evaluation (Sawant and Pfafferott, 2017). Ambient temperature and thermal load profiles were input to the model as look-up tables so as to use the same values which were measured during the experiment. The load is connected to layer 6 and that is applied as the value for the tank model parameter “Load_layer”. The data was logged with a change-of-value protocol and the logging dead-band was 2 % of previous value. The logged data was then extracted in a 60 seconds interval.

For the visual analysis two types of graphical representations are used. The first type shows the measured values and estimated or simulated values over time. This representation expresses the results in an engineering context and helps better understand the physical interactions between the components, which is an important aspect when the application of the models is for system wide simulation. For example, the total duration of the test before the tank set-point temperature is achieved. It also helps to analyse important behaviours of the components like portraying internal control logic and dynamic or quasi-static behaviour. *Fig. 7* to *Fig. 10* show this representation for one test selected per operational mode. The second type shows measured values versus estimated values for outputs of the model that are relevant in terms of control of the system. *Fig. 11* shows this representation for the same selected tests. In both graphs, solid lines denote measured values and dashed lines denote estimated values from simulation.

In *Fig. 7*, the Summer Electricity Production (SEP) mode is simulated. Here, the excess heat from the CHP is stored in the HTES and is used to drive the AdCM and cool down the CTES. Similar to the experiment, a homogeneous initial temperature of 60.3 °C in the HTES and 16.6 °C in the CTES was used. The volume flows in the HT, MT and LT circuits were 1.3 m³/h, 4.2 m³/h and 1.7 m³/h respectively. In addition, a control signal of 1.5 V was applied to the OC and the volume flow in the OC circuit was 4.7 m³/h. The AdCM is switched on at time = 0 minutes.

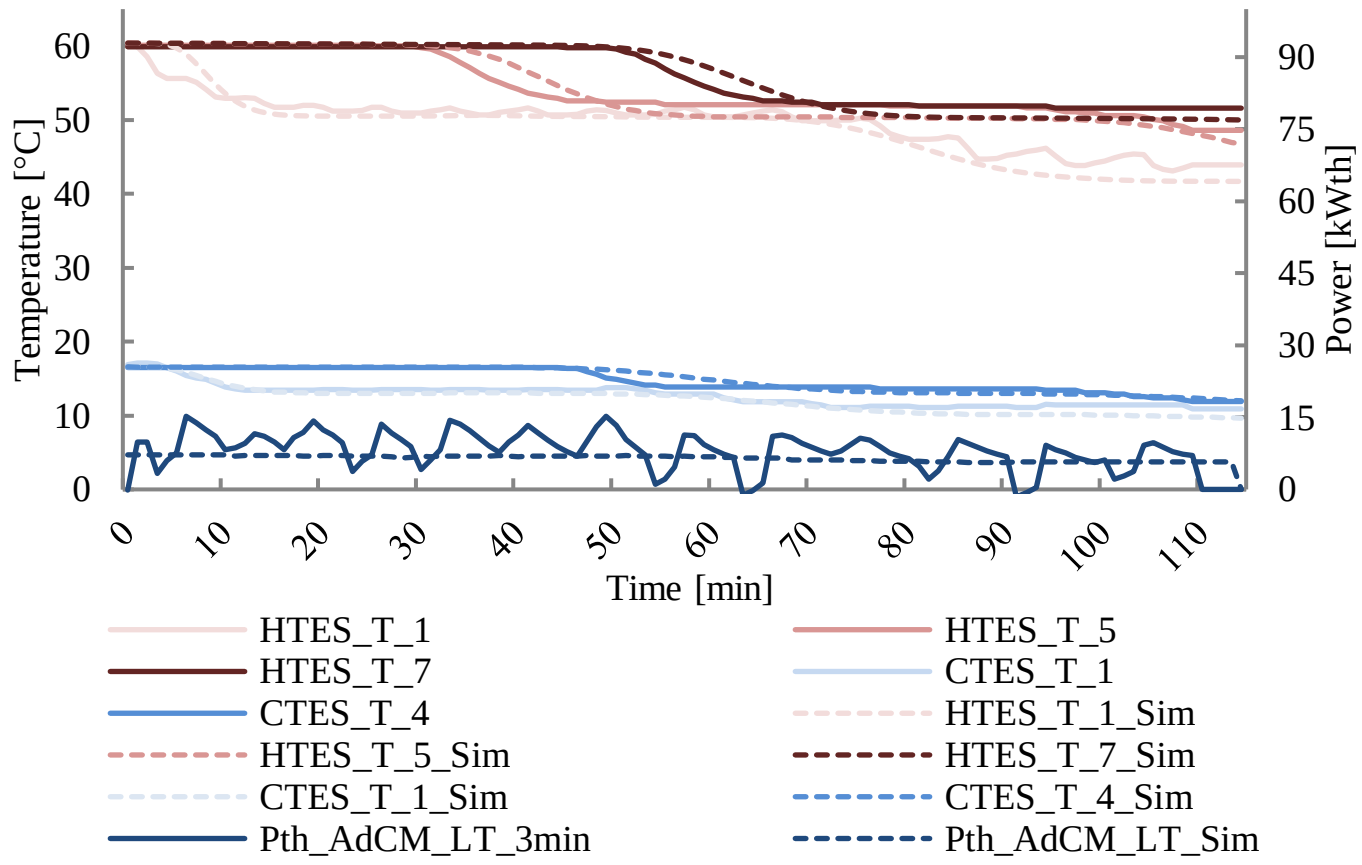

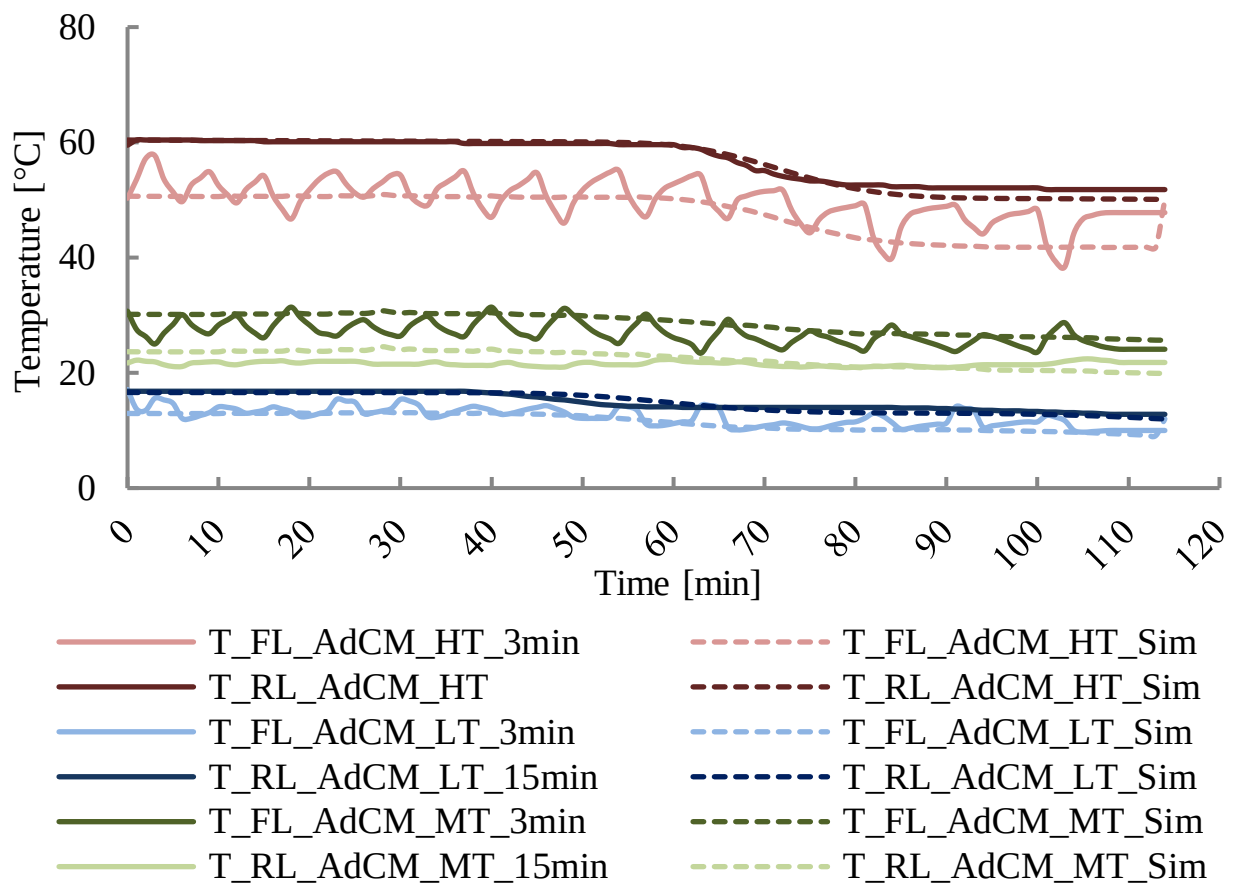

Fig. 7 Experimental and simulation data for the AdCM mode. (a) Tank temperatures and cooling capacity. (b) AdCM circuit temperatures.

To filter the noise in the measured data due to the periodic behaviour of the AdCM, a 3-minute average of the circuit temperatures and cooling capacity $P_{th_AdCM_LT}$ is utilised. Two out of four temperatures in the CTES (CTES_T_1 & CTES_T_4) with CTES_T_1 at the bottom and CTES_T_4 at the top are shown in *Fig. 7 (a)*. Also, three out of the nine temperatures in the HTES (HTES_T_1, HTES_T_5 & HTES_T_7), with HTES_T_1 being at the bottom of the tank and HTES_T_7 corresponding to the outlet to the AdCM, are shown. A visual comparison shows a good fit for the tank temperatures with a deviation in the range 1 to 4 K. The cooling down stratification behaviour is simulated in the cold tank as in the real case and the cyclic behaviour from the circuit temperatures shown in *Fig. 7 (b)* is noticed to be damped in the tank temperatures. The static model simulates cooling power $P_{th_AdCM_LT_Sim}$ as 8 $kW_{th}$ from the beginning of the simulation whereas $P_{th_AdCM_LT}$ has a delay time of 2 minutes before registering the first change. The periodic behaviour of the circuit temperatures and correspondingly the thermal power is not observed in simulation results. Another characteristic simulated is the decrease in cooling as the tank temperatures and correspondingly return-line temperatures in the HT and LT circuits decrease. The experiment lasted 110 minutes compared to the 114 minutes in simulation to achieve the set-point temperature of 12 °C for CTES_T_4.

In *Fig. 8*, the Winter Electricity Production (WEP) mode is simulated. Here the heat from CHP is stored in the HTES and is used to cover the loads. Similar to the experiment a homogeneous initial temperature of 43°C in the HTES was used. $P_{el_CHP_Nom}$ and $P_{th_CHP_Nom}$ are 5 $kW_{el}$ and 10.2 $kW_{th}$ respectively. $\eta_{el_Nom}$ and $\eta_{th_Nom}$ are 0.24 and 0.65 respectively, and the $HCV_{Fuel}$ is taken as 12 kWh/m³ assuming a gas CHP (Bundesnetzagentur, 2019; SenerTech GmbH, 2014). The CHP is switched on at time = 0 minutes. Three out of nine temperatures in the HTES (HTES_T_1, HTES_T_5 & HTES_T_9) with HTES_T_1 being at the bottom of the tank are shown in *Fig. 8 (a)*. A visual comparison shows temperature deviation in the range 1 to 6 K in the HTES temperatures. Thermal stratification behaviour is observed both in the experimental and simulation results. The main outputs of the CHP model are the feedline temperature leaving the CHP $T_{fl_CHP}$ and the volume flow of water $\dot{v}_{CHP}$ which is controlled by the internal controller of the CHP to achieve a maximum possible feedline temperature *Fig. 8 (b)*. Visual analysis shows good accuracy for both outputs in the steady state. The dynamic behaviour of the CHP's thermal power $P_{th_CHP}$ during the start-up phase is also observed with a deviation of around 1 $kW_{th}$ for the first 60 minutes and then a better fit is noticed in steady state. The electric power $P_{el_CHP_Sim}$ shows a static response whereas $P_{el_CHP}$ displays a fast dynamics behaviour with a relatively short time constant of approx. 4.5 minutes. In the experiment the CHP turns off after 446 minutes and in the simulation after 453 minutes once HTES_T_1 reaches a set-point temperature of 72°C. Although the thermal power and volume flow do not turn zero due to the dynamic equations, the formulation in (5) ensures that no mass flow occurs when the CHP is turned off and hence the HTES is not affected.

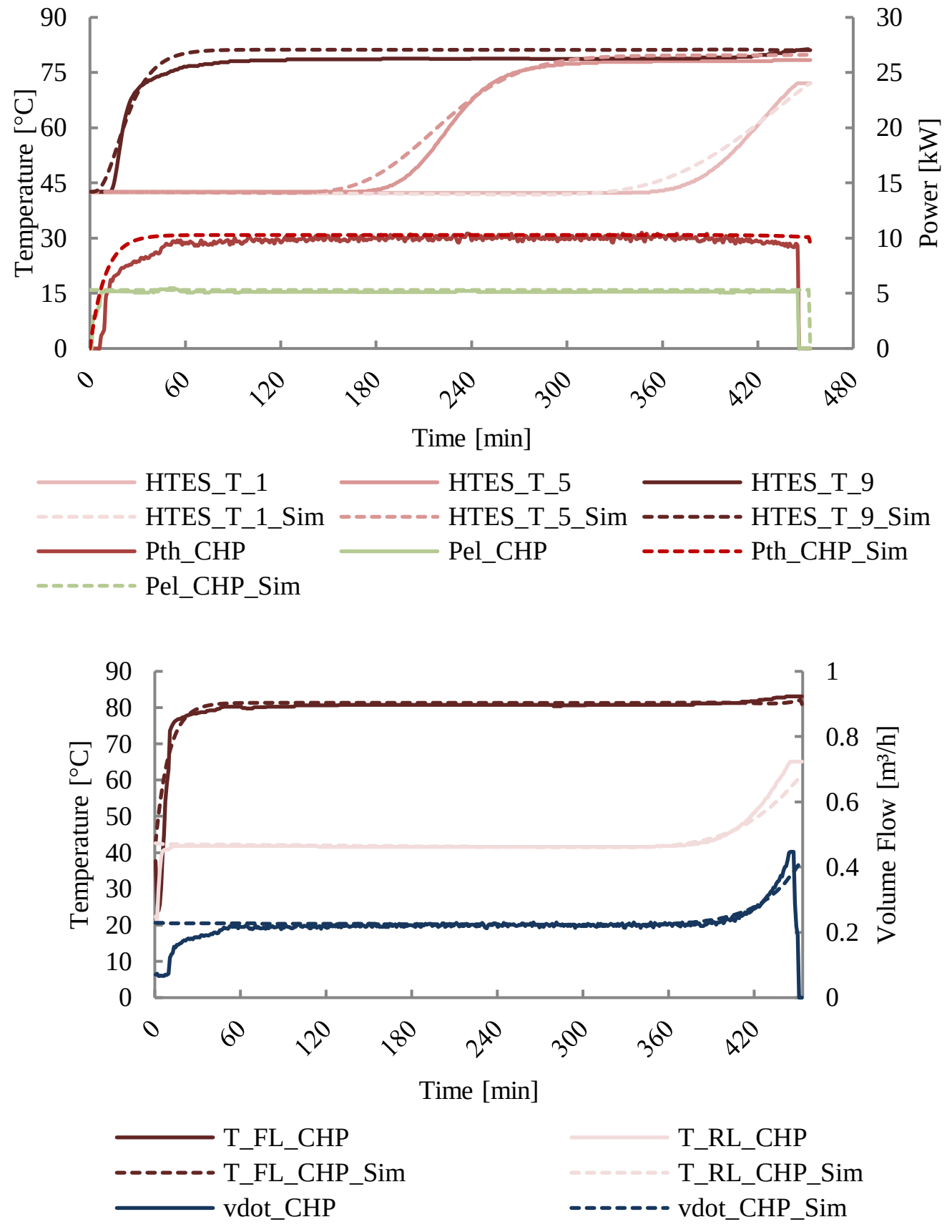

Fig. 8 Experimental and simulation results for the WEP mode. (a) HTES temperatures and CHP powers. (b) CHP circuit temperatures and volume flow.

In *Fig. 9* the Summer Electricity Consumption (SEC) mode is simulated. Here, the cooling capacity of the CCM $P_{th_CCM_LT}$ cools the CTES down. Similar to the experiment a homogeneous initial temperature of 28 °C in the CTES was used. The volume flows in the MT and LT circuits were 2.65 m³/h and 2.45 m³/h respectively. In addition, a control signal of 10 V was applied to the OC. The CCM is switched on at time = 0 minutes. The four temperatures in the CTES (CTES_T_1 to CTES_T_4 with CTES_T_1 at bottom) are shown in *Fig. 9 (a)*. The cooling down stratification behaviour is simulated in the cold tank as in the real case. The main outputs of the CCM model are the two feedline temperatures in the medium and low temperature circuits of the machine $T_{fl_\mathrm{CCM}_MT}$ and $T_{fl_\mathrm{CCM}_LT}$ respectively, the cooling capacity $P_{th_\mathrm{CCM_LT}}$ and the electric consumption $P_{\mathrm{el_RevHP}}$ as shown in *Fig. 9 (a & b)*. Additionally, the circuit temperatures of the OC model are shown. The OC receives a relatively steady 35 °C in its return line and cools it down to almost the ambient temperature $T_{Amb}$. This is in accordance to the fact that the OC is operating at its maximum speed due to the 10 V signal. A visual comparison shows temperature deviation in the range of 1 to 2 K in the two circuits and 1 to 4 K in the tank temperatures. The cooling capacity and electric power consumption deviate by less than 1 kW from the measured values. The $P_{th_\mathrm{CCM_LT_Sim}}$ and $P_{\mathrm{el_RevHP_Sim}}$ display static behaviours and their real values display quasi-static behaviours with a relatively short delay time of approx. 1 minute. Another characteristic simulated is the decrease in cooling as the CTES temperatures and correspondingly the return-line temperature in the LT circuit decreases. In the experiment the machine ran for 158 minutes and in simulation for 163 minutes before turning off due to achieving set temperature of 10 °C for CTES_T_4.

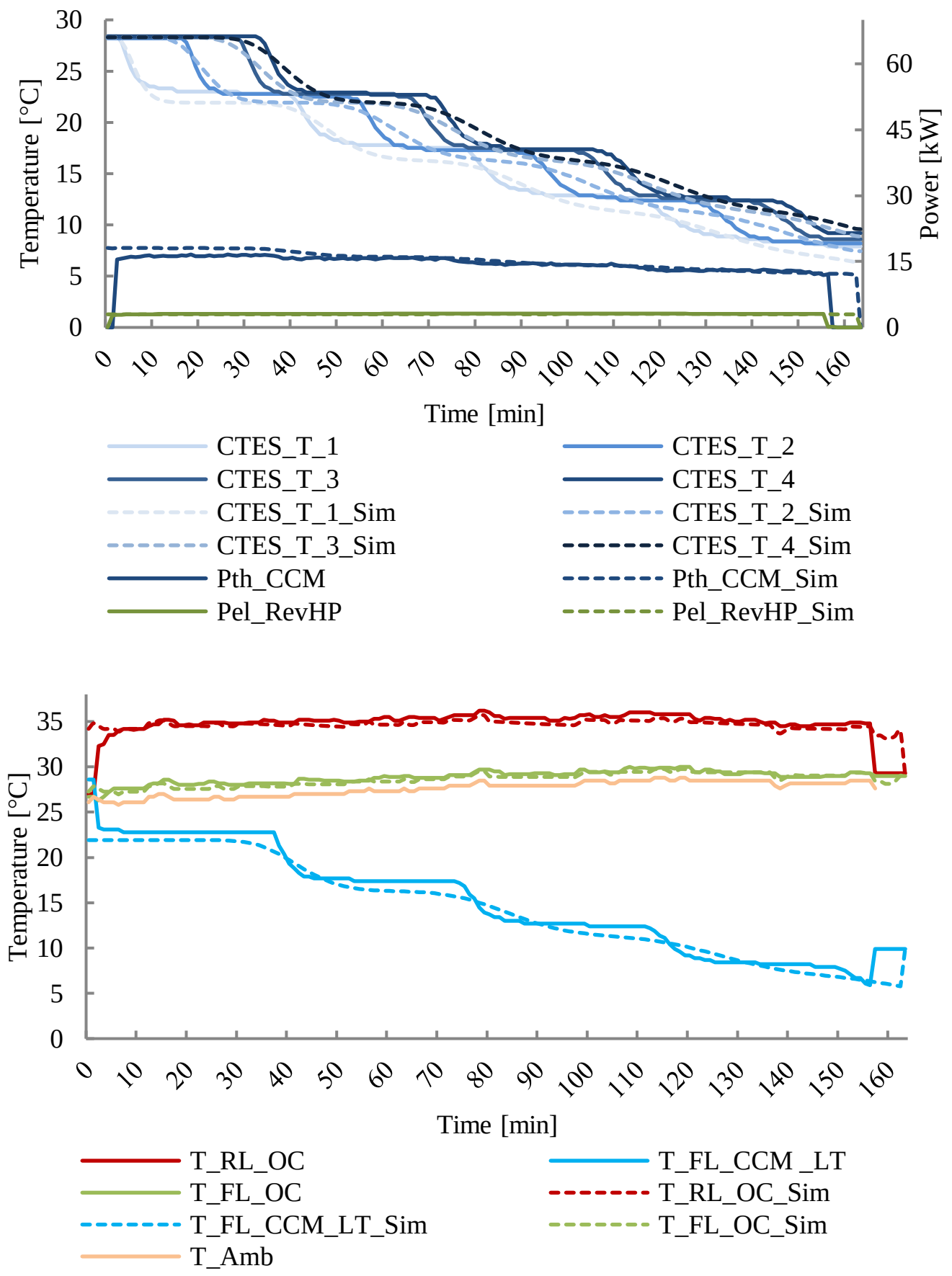

Fig. 9 Experimental and simulation results for the SEC mode. (a) CTES temperatures and CCM powers. (b) CCM and OC circuit temperatures.

In *Fig. 10*, the Winter Electricity Consumption (WEC) mode is simulated. Here the heating capacity of the HP $P_{th_\mathrm{HP_HT}}$ heats the HTES to satisfy thermal loads. Similar to the experiment a homogeneous initial temperature of 20 °C in the HTES was used. The volume flows in the HT and MT circuits were 1.0 m³/h and 2.4 m³/h respectively. The volume flow in the HT circuit was measured using an ultrasonic flow meter since the value is not measured continuously in the real plant. In addition, a control signal of 10 V was applied to the OC and the volume flow in the OC circuit was 4.7 m³/h. The AdCM is switched on at time = 0 minutes. Three out of nine temperatures in the HTES (HTES_T_1, HTES_T_5 & HTES_T_9) are shown in *Fig. 10 (a)* with HTES_T_1 being at the bottom of the tank. A visual comparison shows an error in the range of 1 to 6 K in the tank temperatures. A limited thermal stratification behaviour is observed both in the experimental and simulation results due to a smaller temperature difference in the circuits of the HP. The main outputs of the HP model are the feedline temperatures leaving the HP circuits $T_{fl_\mathrm{HP}_MT}$ and $T_{fl_\mathrm{HP}_HT}$, the heating capacity $P_{th_\mathrm{HP_HT}}$ and the electric consumption $P_{\mathrm{el_RevHP}}$ shown in *Fig. 10 (a & b)*. A deviation of less than 2 K is seen in the MT circuit and a deviation of around 6 K is seen in the HT circuit. Like in the case of the CCM, the powers in the HP simulation $P_{th_\mathrm{HP_HT_sim}}$ and $P_{\mathrm{el_RevHP_sim}}$, also display static behaviours and their real values display quasi-static behaviours with a relatively short delay time of approx. 6 minutes. In the experiment, the HP turns off after 165 minutes and in the simulation after 166 minutes when HTES_T_1 reaches its set-point temperature of 40 °C.

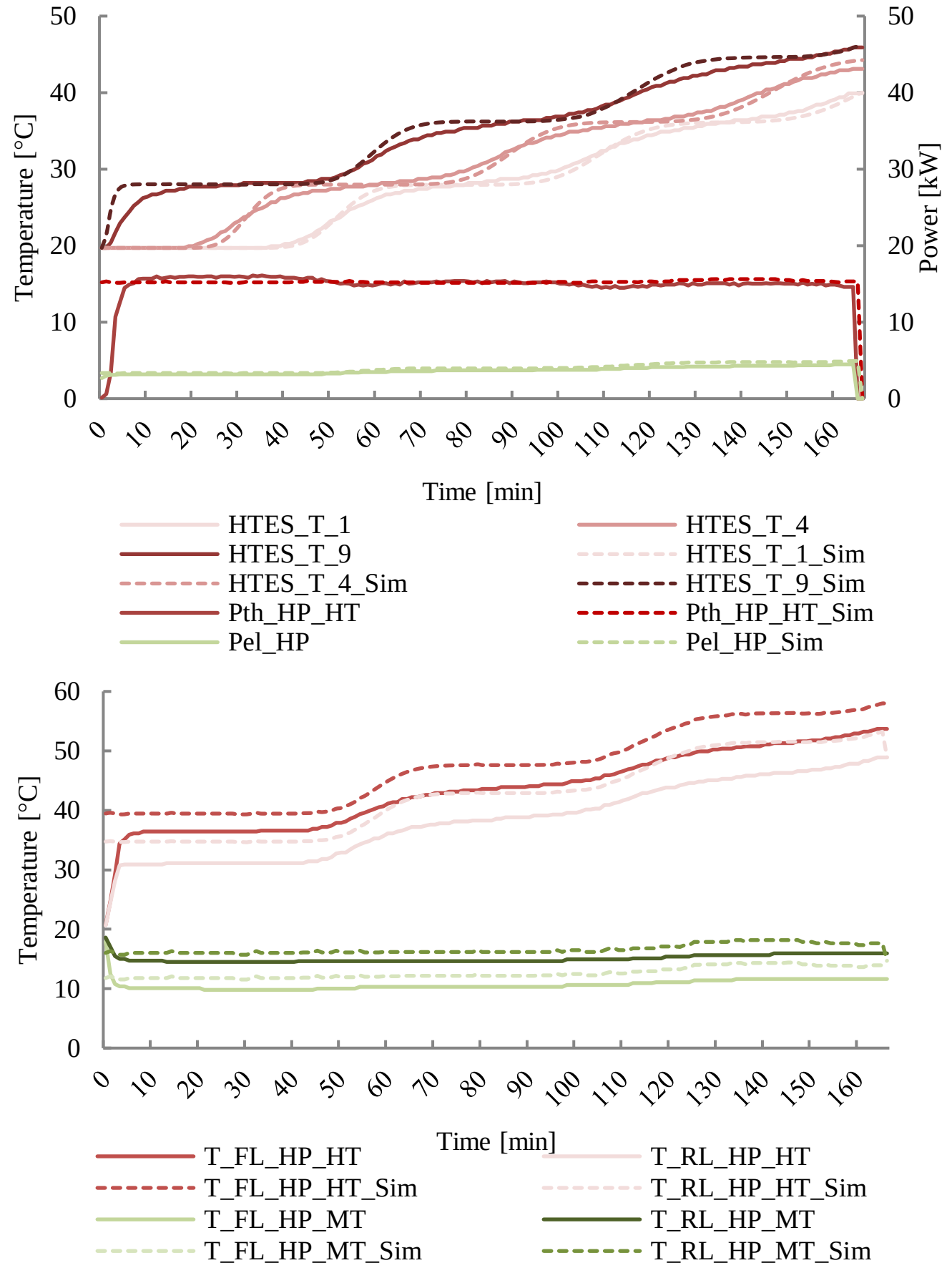

Fig. 10 Experimental and simulation results for the WEC mode. (a) HTES temperatures and HP powers. (b) CHP circuit temperatures.

In *Fig. 11*, the quality of the results are illustrated in more detail by comparing the measured and estimated tank temperature that is relevant in the case of the particular mode of operation. This facilitates the analysis of the simulation of the complex physical interactions in an energy system because the tanks are the hydraulic and thermal interface between the source and load sides.

In *Fig. 11* (a), the CTES_T_1 (bottom of CTES going to Load_C) temperature is shown for the SEP mode. The maximum deviation is 15 % of measured value and amounts to 1.7 K when CTES temperature is approx. 11 °C. The stratification behaviour is observed in this representation also. The model often over-estimates the CTES temperature but within a range of 5 % of measured value.

In *Fig. 11* (b), the HTES_T_6 (middle of HTES going to Load_H) is shown for the WEP mode. The maximum deviation is when the model over-estimates the temperature at lower tank temperatures between 42 to 50 °C. This amounts to around 6 K difference. The stratification behaviour is also observed in this representation.

In *Fig. 11* (c), the CTES_T_1 temperature is shown for the SEC mode. The maximum deviation is 15 % of measured value and amounts to approx. 1.5 K. Cooling down stratification is observed. Correspondingly the model mostly over-estimates the CTES temperature but within a 15 % range.

In *Fig. 11* (d), the HTES_T_6 is shown for the WEC mode. The maximum deviation is 8 % of measured value and thermal stratification is also simulated.

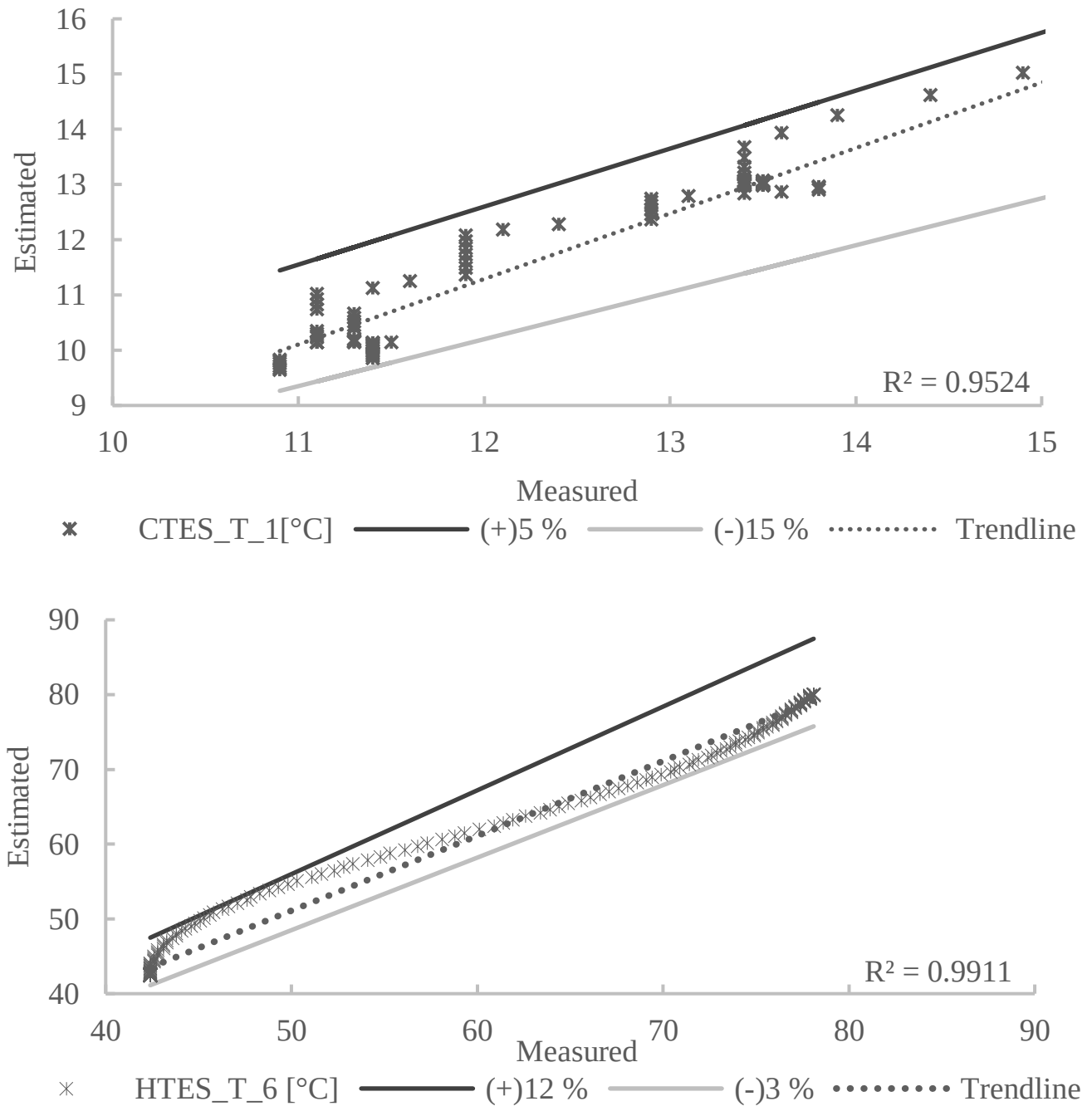

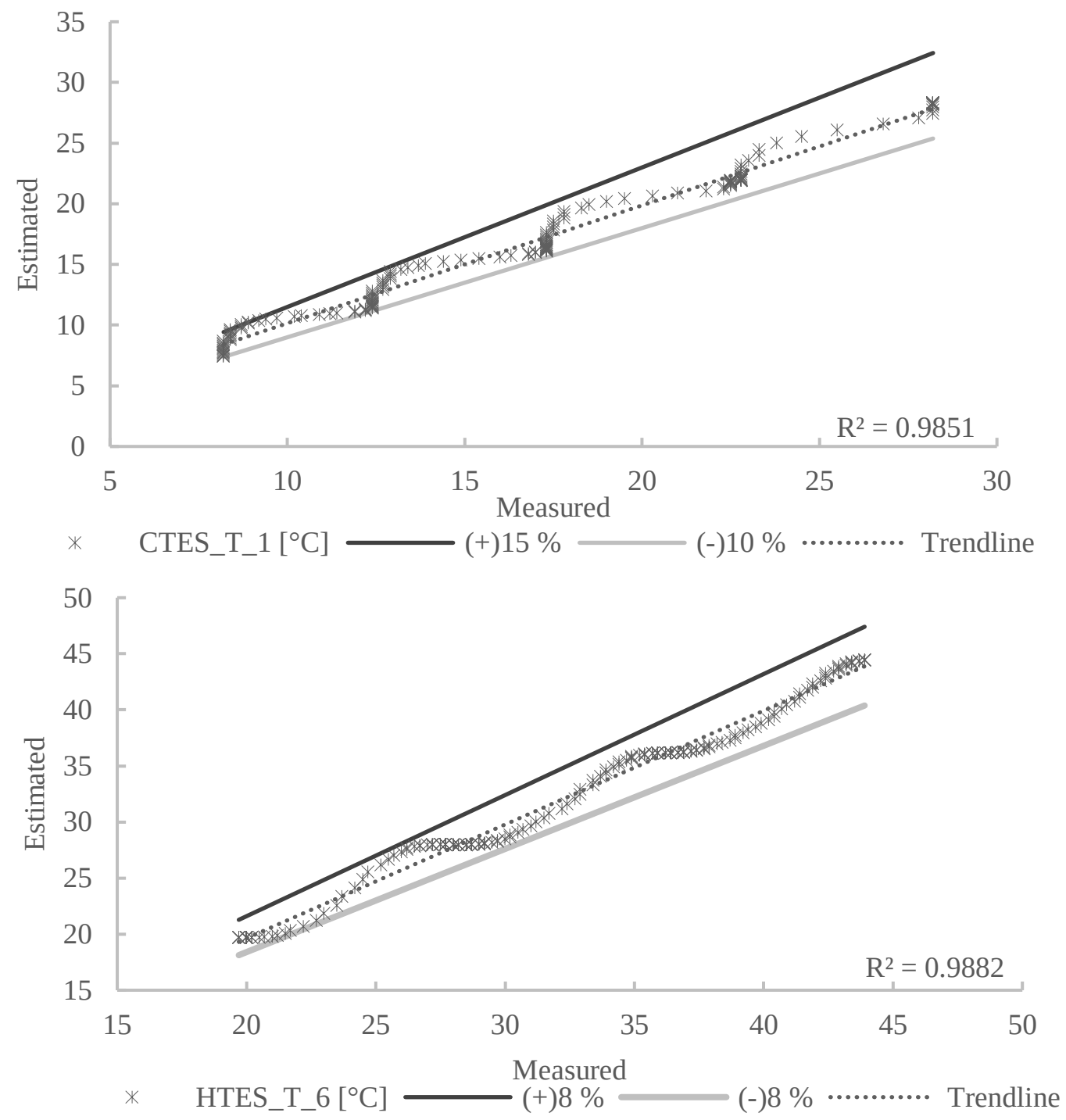

Fig. 11 Measured versus estimated results for the relevant tank temperatures.

For the quantitative analysis the following standard metrics were chosen amongst the most commonly used statistical methods in the HVAC simulation world (Afram and Janabi-Sharifi, 2015a):

$$NRMSRE = \sqrt{\frac{\sum_{i=1}^{n}\left(\frac{y_i - y_i^*}{y_{max} - y_{min}}\right)^2}{n}} \tag{46}$$

$$r^2 = \left(\frac{\sum_{i=1}^{n}(y_i - \bar{y})(y_i^* - \bar{y}^*)}{\sqrt{\sum_{i=1}^{n}(y_i - \bar{y})^2 \sum_{i=1}^{n}\left(y_i^* - \bar{y}^*\right)^2}}\right)^2 \tag{47}$$

$$GoF = 100\left(1 - \frac{\sqrt{\sum_{i=1}^{n}\left(y_i^* - y_i\right)^2}}{\sqrt{\sum_{i=1}^{n}(y_i - \bar{y})^2}}\right) \tag{48}$$

where,
$NRMSRE$ – Normalised root mean squared relative error
$r^2$- Square of the Pearson product moment correlation coefficient (Microsoft Corporation, 2019)
$GoF$- Goodness of fit
$y_i$- i[th] measured value
$y_i^*$- i[th] predicted value
$\bar{y}$- Mean of measured values data set
$\bar{y}^*$- Mean of predicted values data set
$y_{max}$- Maximum value of $y$ in entire data set
$y_{min}$- Minimum value of $y$ in entire data set
$n$ – Number of data points

The closer the value for *NRMSRE* is to zero the better is the fit whereas the closer the value for $r^2$ and $GoF$/100 is to 1.0 the better is the fit.

Data from three to four functional tests per operational mode was accumulated and the evaluation metrics for some of the main model outputs were calculated as shown in *Fig. 12* to *Fig. 15*. The duration of the tests was between 2 to 8 hours depending on initial tank temperatures or thermal loads.

In *Fig. 12*, the results of the AdCM model are shown. The values for CTES_T_1 show best fits with $NRMSRE$ of 0.11, $r^2$ of 0.88 and $GoF$ of 52.5. The $P_{th_AdCM_LT}$, $COP$, and $T_{fl_\mathrm{AdCM_LT}}$ have a $NRMSRE$ higher than 0.15 and a $r^2$ less than 0.4. The $GoF$ for $P_{th_AdCM_LT}$ and $COP$ is negative.

In *Fig. 13*, the results of the CHP model are shown. The values for HTES_T_6 and $T_{fl_\mathrm{CHP}}$ show good fits with $NRMSRE$ of 0.13, $r^2$ > 0.90 and $GoF$ > 58.0. The $P_{th_\mathrm{CHP}}$ and $\dot{v}_{CHP}$ have a *NRMSRE* higher than 0.2 and a $r^2$ less than 0.55. The $GoF$ for both is also less than 20.

In the *Fig. 14* the results of the CCM model are shown. The values for CTES_T_1 and $T_{fl_\mathrm{CCM_LT}}$ show good fits with $NRMSRE$ of < 0.05, $r^2$ > 0.96, and $GoF$ > 77.0. The $P_{th_\mathrm{CCM}}$ and $COP$ have a $NRMSRE$ higher than 0.19 and an $r^2$ less than 0.4. The $GoF$ for the *COP* is 88.0. However, the $GoF$ for $P_{th_\mathrm{CCM}}$ is only 10.5.

In the *Fig. 15*, results of the HP model are shown. The values for HTES_T_6 show best fits with $NRMSRE$ of 0.05, $r^2$ of 0.96 and $GoF$ of 76.0. The $P_{th_\mathrm{HP}}$ and $P_{el_\mathrm{RevHP}}$ have an $NRMSRE$ higher than 0.19 and an $r^2$ less than 0.4. The $GoF$ for $P_{th_\mathrm{HP}}$ and $T_{fl_\mathrm{HP_HT}}$ is less than 20.0. However, the $GoF$ for $P_{el_\mathrm{RevHP}}$ is high with a value of 85.0.

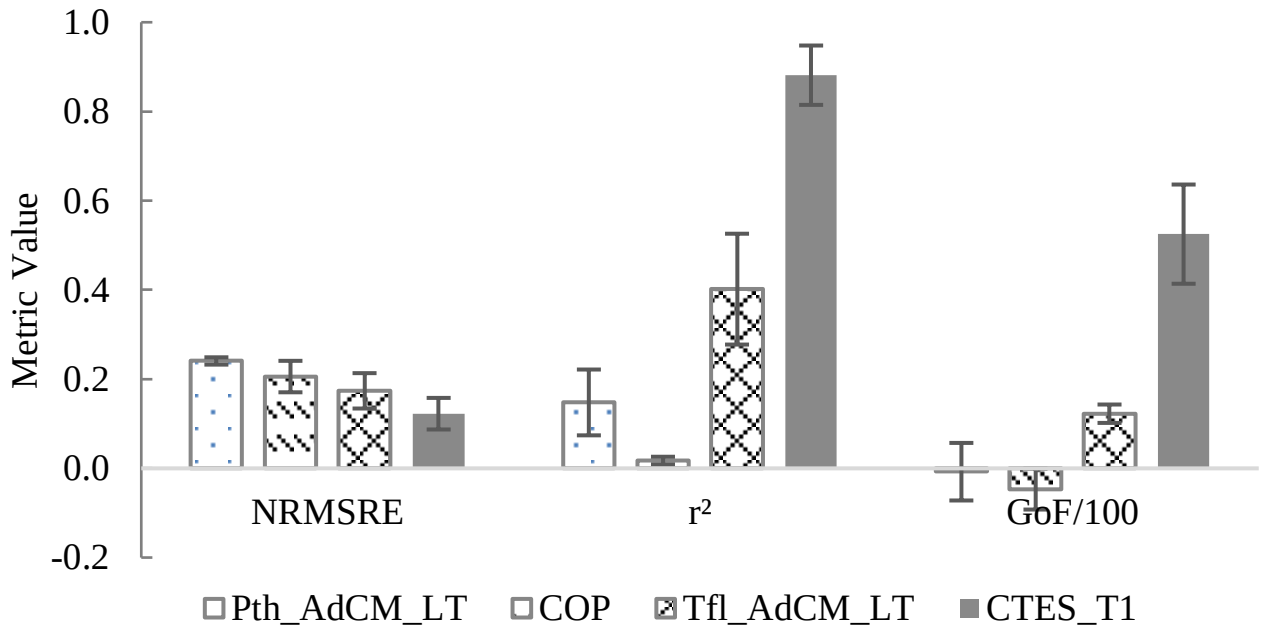

Fig. 12 Evaluation metrics for the AdCM outputs.

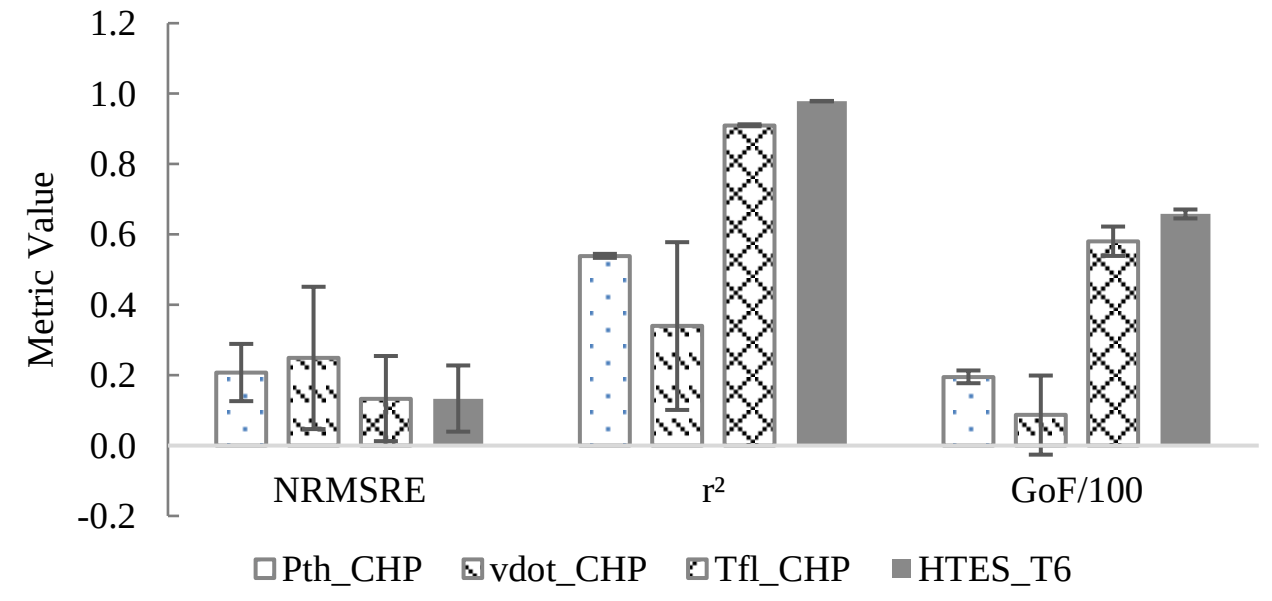

Fig. 13 Evaluation metrics for the CHP outputs.

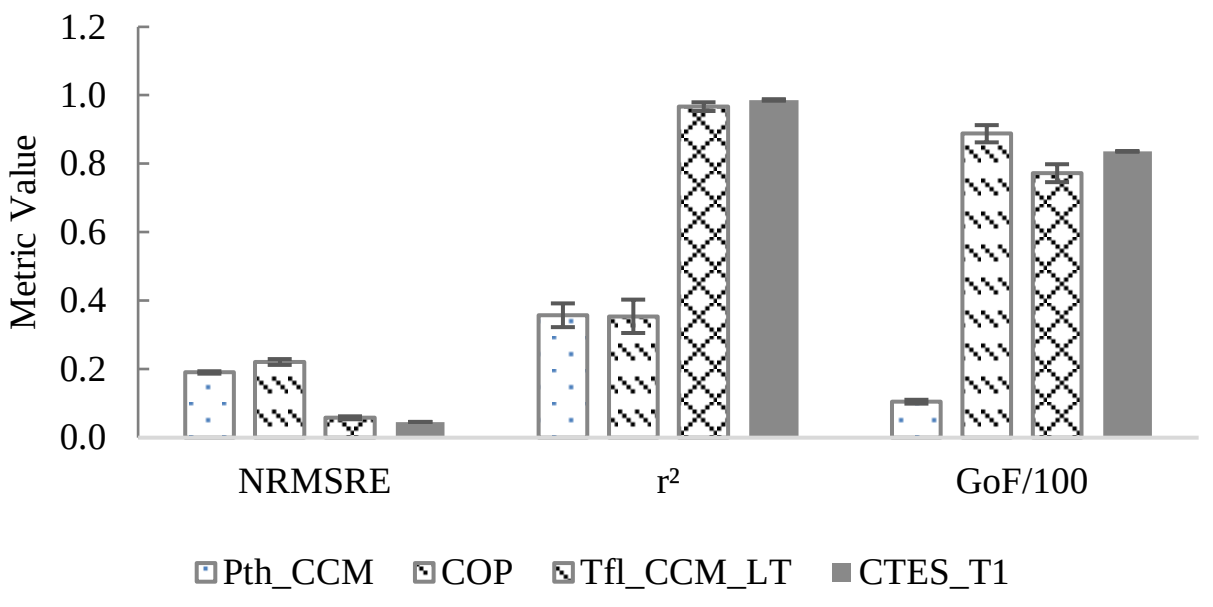

Fig. 14 Evaluation metrics for the CCM outputs.

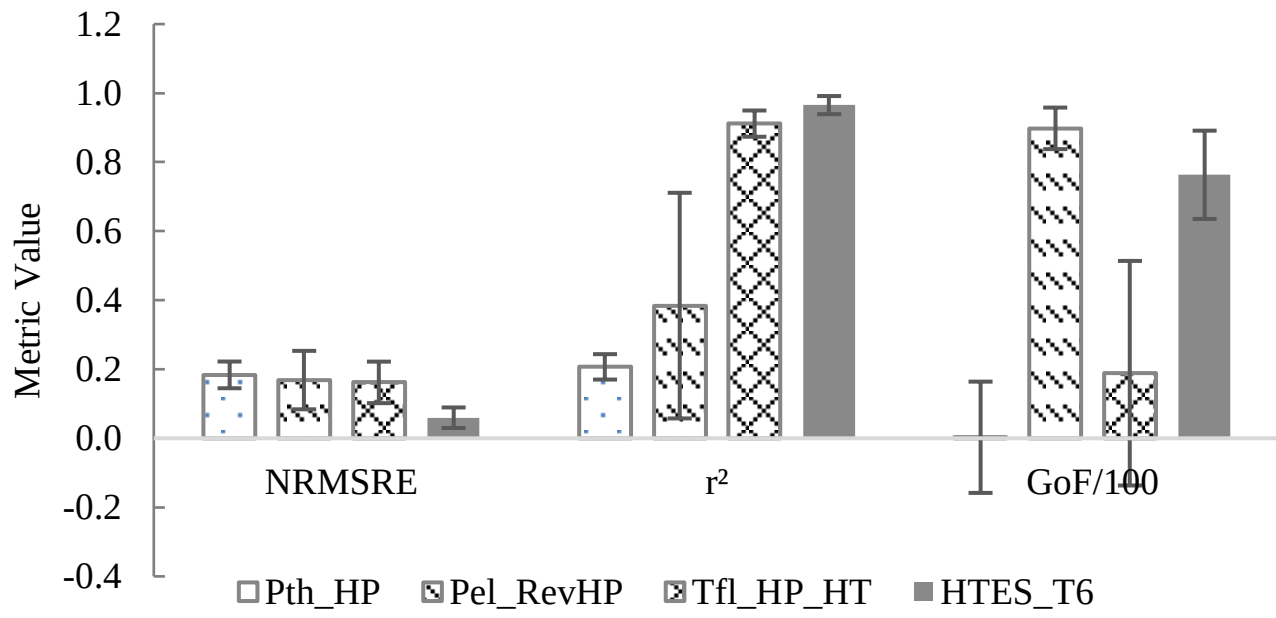

Fig. 15 Evaluation metrics for the HP outputs.

## 5. DISCUSSION & CONCLUSIONS

Based on an extensive literature research, a rational modelling approach to develop component-models of a complex polygeneration system with reduced input parameters and limited empirical data was exhibited in a realistic scenario. The models were visually and quantitatively evaluated for their application in a supervisory MPC.

From the visual analysis of the AdCM, HTES and CTES model in *Fig. 7* a good fit for the tank temperatures is seen, which is specifically highlighted in the *Fig. 11* (a) and thus demonstrates the good ability of this model to take advantage of a system wide simulation. However, our model does not simulate the periodic behaviour of the circuit temperatures and the start-up phase of the AdCM, which reflects in the poor performance in the quantitative analysis as seen in *Fig. 12*. More detailed models in literature are capable of simulating this behaviour and have a better performance (*Table 3*). For example some report an average percentage error of 5.6 % for the cooling capacity and 4.1 % for the COP (Li and Wu, 2009) compared to 21 % and 16 % for our model. However, as mentioned earlier these models are extremely complex and not suitable for the system-wide optimal control approach, which is the focus in our work. The AdCM model in this paper compromises on accuracy and is not sufficient for AdCM specific analysis, but due to its simplicity it is suitable for optimisation-based control of an energy system comprising of AdCM and storages.

From the visual analysis of the CHP and HTES model in *Fig. 8* a good fit for the tank temperatures and the powers of the CHP is noticed. The HTES temperature is specifically highlighted in *Fig. 11* (b) and thus the ability of the two models to simulate the system dynamics with high accuracy is established. The model is capable of producing good results even for the main outputs of the CHP itself and this is reflected in the quantitative analysis as seen in *Fig. 13*. The evaluation metrics for $P_{th_\mathrm{CHP}}$ and $\dot{v}_{CHP}$ could be improved further by using more high quality data for fitting the coefficients. Optimal experimental design techniques and more steady state data at different $T_{rl_\mathrm{CHP}}$ could be used for example. Our model contributes to the existing literature by combining certain important approaches and using lesser parameters to produce similar visually analysed results (Chandan et al., 2012; Seifert, 2013). The CHP model is suitable for application in system-wide optimisation and energy system analysis.

From the visual analysis of the RevHP, OC and tank models in *Fig. 9* and *Fig. 10* a good fit for the tank temperatures and the powers of the RevHP is observed. The tank temperatures are specifically highlighted in *Fig. 11* (c & d) and thus the ability of the models to simulate the complex interactions with high accuracy is established. The model is capable of producing good results even for the main outputs of the RevHP itself and this is reflected in the quantitative analysis as seen in *Fig. 14* and *Fig. 15*. The simulation results for the HT circuit temperatures of the HP as seen in *Fig. 10 (b)* could be improved further by doing a more accurate parameterisation of the heat exchanger between the HP and the HTES and using a more accurate measurement method for the volume flow in this circuit. The experimental validation of these models reinforces the validity of the approaches presented in published scientific material. For example, the absolute error for $T_{fl_\mathrm{AdCM_LT}}$ is less than 10 % (Yao et al., 2013) and the mean absolute percentage error for $P_{th_\mathrm{HP_HT}}$ is approx. 5 % (Salvalai, 2012).

From the quantitative analysis using different evaluation metrics it is established that no particular metric is suitable for the evaluation of all the variables and the quantitative analysis should be performed in the context of the visual analysis. This reiterates from other works where it is mentioned to use different metrics to assess the quality of the models based on the developer's criteria (Fumo and Rafe Biswas, 2015). We used the $NRMSRE$, $r^2$ and $GoF$ methods. The $NRMSRE$ shows stable results and a value of less than 0.15 is considered as a good fit in this work. On the other hand, the $r^2$ and $GoF$ are extremely sensitive to the errors caused from mismatch of time-series or when neglecting dynamics of components. For example, in the WEC mode in *Fig. 10* when the simulation lasts 1 minute longer than the experiment and the first 7 minutes of $P_{th_\mathrm{HP_HT}}$ dynamics are not simulated accordingly, the $GoF$ is very low. Similar is the case for variables like COPs and volume flows which vary suddenly due to static simulations. However, for tank temperatures and circuit temperatures the $r^2$ and $GoF$ could be a suitable metric.

Further arguments in favour of this modelling approach are as follows:

- Although a deviation of up to 5 K is noticed in the tank temperatures, this is not over an exceptionally longer period of time and the thermal inertia of HVAC systems will alleviate this deviation.
- Static models for AdCM and RevHP are justified since the time constants of these components are typically smaller than 5 minutes. For plant operations having normally two to three start-up/shut-down cycles over an entire day it is not indispensable to model their dynamics.
- These models will be applied in a 15 minute energy-market-price based optimisation problem and thus a deviation of 5

to 10 minutes in total operational time as seen in *Fig. 7* to *Fig. 10* could be considered a very good fit. However, this renders these models unsuitable for grid voltage or frequency management based optimal scheduling problems.

Drawing from the quantitative and qualitative arguments presented above it is concluded that the models presented in this work are of sufficient accuracy and stability to be integrated in an optimisation framework since they are also generalizable and reflect other internal control and part-load aspects of the components.

In our future work, these models will be utilised within a Mixed-Integer Optimal Control Problem formulation for economic-MPC of the system. For solution of such problems, application of direct methods such as direct multiple shooting (Bock and Plitt, 1984) and direct collocation (Tsang et al., 1975) is favourable. This results in Mixed-Integer Nonlinear Programs, which will be implemented using CasADi and solved using real-time suitable methods, e. g., as presented by Sager, 2009 (Sager, 2009). Subsequently we will extend the automation and control architecture (ACS) to fit the coefficients of regression in real-time. There, the efficacy of our modelling approach for implementation in a supervisory-MPC in a retrofit or a green-field scenario will be demonstrated.

## ACKNOWLEDGEMENTS

The authors are grateful to the Reiner Lemoine Stiftung, Alexander von Humboldt Stiftung, the DENE Graduate College (Baden-Württemberg MWK), E-Werk Mittelbaden Innovations Fond and the State Ministry of Baden-Wuerttemberg for Sciences, Research and Arts (Az: 22-7533.-30-20/9/3) for their support.

Conflict of interest - none declared.